\documentclass[11pt]{article}

\usepackage[preprint]{acl}

\usepackage{times}
\usepackage{latexsym}
\usepackage{booktabs}
\usepackage{multicol}
\usepackage{multirow}
\usepackage{hyperref}
\usepackage{url}
\usepackage{comment}

\usepackage{booktabs}  
\usepackage{multirow}
\usepackage{algorithm}
\usepackage{algpseudocode}
\usepackage{caption}
\usepackage{mathtools}
\usepackage[most]{tcolorbox}
\usepackage{listings}
\usepackage{xcolor}

\tcbset{
  promptstyle/.style={
    colback=gray!8,      % 背景灰度更深
    colframe=black!40,   % 边框稍深
    boxrule=0.5pt,
    arc=2pt,
    left=6pt,
    right=6pt,
    top=5pt,
    bottom=5pt,
  }
}

\usepackage[T1]{fontenc}
\usepackage[utf8]{inputenc}

\usepackage{microtype}

\usepackage{inconsolata}

\usepackage{graphicx}
\usepackage{listings}
\usepackage{xcolor}

\lstdefinestyle{promptclean}{
  basicstyle=\ttfamily\footnotesize,
  backgroundcolor=\color{gray!5},
  frame=single,
  breaklines=true,
  columns=fullflexible,
  keepspaces=true,
  tabsize=2,
  showstringspaces=false,
  numbers=none,
  captionpos=none,     % 无标题
  abovecaptionskip=0pt,
  belowcaptionskip=0pt,
  framexleftmargin=1pt,
  framexrightmargin=1pt,
  framextopmargin=2pt,
  framexbottommargin=2pt,
}

\newcommand{\vpara}[1]{\vspace{0.5em}\noindent\textbf{#1}\quad}

\title{WebVIA: A Web-based Vision-Language Agentic Framework for Interactive and Verifiable UI-to-Code Generation}

\author{
Mingde Xu$^{1,3}$\thanks{\ \ Core contributors.}, 
Zhen Yang$^{2,3}$\footnotemark[1]\thanks{\ \ Corresponding author.}, 
Wenyi Hong$^{2,3}$,   
\textbf{Lihang Pan$^3$, Xinyue Fan$^3$,}
\\
\textbf{Yan Wang$^3$, Xiaotao Gu$^3$,} 
\textbf{Bin Xu$^2$,}
\textbf{Jie Tang$^2$}\footnotemark[2] \\
$^1$Faculty of Mathematics, University of Waterloo \\
$^2$The Knowledge Engineering Group (KEG), Tsinghua University\\
$^3$Zhipu AI\\
\small{\texttt{{m339xu@uwaterloo.ca,}}}
\small{\texttt{{yang-zhen@mail.tsinghua.edu.cn,}}}  \small{\texttt{{jietang@mail.tsinghua.edu.cn}}}
\\
}

\begin{document}
\maketitle
\begin{abstract}
User interface (UI) development requires translating design mockups into functional code, a process that remains repetitive and labor-intensive.
While recent Vision–Language Models (VLMs) automate UI-to-Code generation, they generate only static HTML/CSS/JavaScript layouts lacking interactivity.
To address this, we propose WebVIA, the first agentic framework for interactive UI-to-Code generation and validation.
The framework comprises three components: 1) an exploration agent to capture multi-state UI screenshots; 2) a UI2Code model that generates executable interactive code; 3) a validation module that verifies the interactivity.
Experiments demonstrate that WebVIA-Agent achieves more stable and accurate UI exploration than general-purpose agents (e.g., Gemini-2.5-Pro). 
In addition, our fine-tuned WebVIA-UI2Code models exhibit substantial improvements in generating executable and interactive HTML/CSS/JavaScript code, outperforming their base counterparts across both interactive and static UI2Code benchmarks. 
Our code and models are available at 
\href{https://zheny2751-dotcom.github.io/webvia.github.io/}{\texttt{https://webvia.github.io}}.

\end{abstract}

% \section{Introduction}

\section{Introduction}

\begin{figure*}[t!]
    \centering
    \includegraphics[width=1\linewidth]{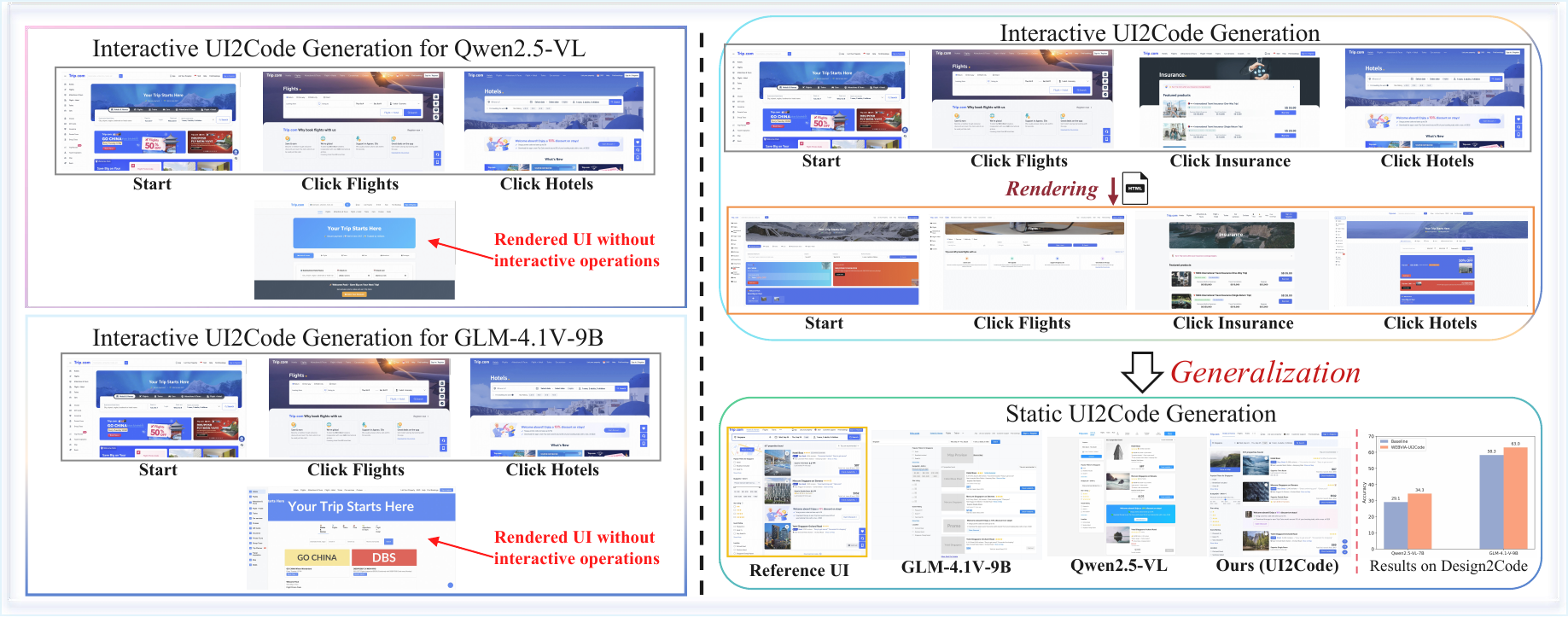}
    % \vspace{-2mm}
    \caption{Motivating example illustrating the gap between static and interactive code generation. }
    \label{fig:motivating_example}
    \vspace{-4mm}
\end{figure*}

User interface (UI) development is a core step in modern software engineering, yet translating design mockups into functional code remains a repetitive and labor-intensive process. Automated UI-to-Code (UI2Code) has therefore emerged as a promising direction, aiming to automatically transform UI screenshots into structured front-end code. Recent advances in Vision-Language Models (VLMs)~\cite{liu2023visual,wang2024qwen2,wang2024cogvlm,bai2025qwen2,hong2025glm,zhu2025internvl3} have created new opportunities for UI2Code that jointly interprets visual layouts and textual semantics, thus moving beyond shallow pattern recognition~\cite{beltramelli2018pix2code,acsirouglu2019automatic,chen2022code} toward more robust and visually grounded code generation~\cite{wu2025mllm,jiang2025screencoder}.

Despite these advances, current VLM-powered UI2Code approaches remain limited in functionality. 
As demonstrated in Figure~\ref{fig:motivating_example}, 
their outputs are typically restricted to static HTML/CSS/JavaScript code, which reproduce the visual appearance of interfaces but lack support for GUI interactions such as clicking, selecting, or entering text. 
The generated interfaces cannot correctly respond to user actions and therefore cannot be integrated into real-world UI development workflows. These limitations highlights the necessity for a new framework capable of producing executable and truly interactive user interfaces.

To address this gap, we propose WebVIA, the first agentic framework for interactive UI-to-Code generation and validation. The framework is composed of three components: (1) an exploration agent that interacts with the HTML environment to capture multiple UI screenshots across different interface states; (2) a UI2Code model that leverages these screenshots to generate executable HTML/CSS code for an interactive GUI; (3) a validation module that assesses the feasibility and interactivity of the generated GUI. 

We train two core models to ensure the high performance of WebVIA.
The first is an exploration agent, WebVIA-Agent, that traverses the HTML environment to collect diverse interface states. We construct a large-scale GUI interaction dataset, and experimental results show that WebVIA-Agent outperforms general-purpose models such as Gemini-2.5-Pro~\cite{comanici2025gemini} in both stability and accuracy.
The second is a UI2Code model that generates executable and interactive interfaces. Based on Qwen-2.5-VL-7B and GLM-4.1V-9B, we train WebVIA-UI2Code-Qwen and WebVIA-UI2Code-GLM using paired multi-state UI screenshots and executable interactive HTML. Compared with their respective base models, the two variants achieve improvements of 5.2 and 4.7 points on the Design2Code benchmark, and on UIFlow2Code they reach performance levels (75.9 and 84.9, respectively) close to that of the large-parameter state-of-the-art Claude-Sonnet-4 .

Our contributions can be summarized as follows:

\begin{itemize}
    \item \textbf{Framework:} We propose WebVIA, the first agentic framework for interactive UI-to-Code generation and validation, which bridges the gap between static UI rendering and executable, interactive front-end development.

    \item \textbf{Model:} Under the guidance of WebVIA, we train two dedicated models: an exploration agent for HTML environment interaction and state collection, and a UI2Code model capable of generating executable code that supports real user interactions.

    \item \textbf{Experiments:} In our experiments, the WebVIA-Agent achieves higher stability and accuracy than general-purpose models such as Gemini-2.5-Pro, while the WebVIA-UI2Code model surpasses static layout reconstruction to produce robust and verifiable interactive web code.
\end{itemize}

\section{Related Work}\label{related_work}

\subsection{UI-to-Code Generation}

The UI-to-Code (UI2Code) task aims to translate user interfaces into executable code, evolving from early deep learning approaches to recent vision–language model (VLM)-based methods.
Before VLMs, works like Pix2Code~\cite{beltramelli2018pix2code} used CNN–RNN architectures for static UI generation.
With VLMs, methods such as DeclarUI~\cite{zhou2024bridging} integrate vision and language models for better component grounding, while others enhance performance through large-scale datasets~\cite{Laurencon2024websight,yun2024web2code,gui2024webcode2m,gui2024vision2ui}, multi-model collaboration~\cite{jiang2025screencoder,liang2024waffle}, or fine-grained interface decomposition~\cite{wu2025mllm,chen2025designcoder,wan2024automatically}.
However, these approaches remain limited to static HTML/CSS generation, lacking interactivity or functional validation.
In contrast, WebVIA introduces an agentic, interaction-aware framework that enables verifiable and executable UI2Code generation.

\subsection{Interactive Web Agents}

Recent research has moved beyond static UI translation to explore interactive web environments, where agents perform multi-step actions on real webpages.
Early systems such as World of Bits~\cite{shi2017world}, WebGPT~\cite{nakano2021webgpt}, and BrowserGym~\cite{chezelles2024browsergym} integrate language models with browser environments for reasoning and decision-making over dynamic content.
Later benchmarks like WebArena~\cite{zhou2023webarena} and Mind2Web~\cite{deng2023mind2web} enable grounded web interaction by allowing agents to perceive, plan, and execute GUI actions from both DOM structures and visual inputs.
While these frameworks highlight the value of multimodal reasoning and feedback, existing agents remain optimized for single, predefined tasks, lacking the ability to systematically explore or verify interactive components.
In contrast, our approach extends this paradigm toward agent-based UI synthesis and verification, where the agent actively traverses and tests webpage components to ensure functional and behavioral correctness.

\subsection{UI2Code Benchmark}

Several benchmarks have advanced VLM-based UI2Code research.
Design2Code~\cite{si2024design2Code} introduces 484 real-world webpages for visual-to-code evaluation, while Web2Code~\cite{yun2024web2code} and Flame-React~\cite{ge2025advancing} refine data pipelines via LLM-assisted layout–code synthesis, though still relying on synthetic HTML.
FullFront~\cite{sun2025fullfront} broadens evaluation to the full front-end workflow—design, perception, and code generation.
More recently, Interaction2Code~\cite{xiao2024interaction2code} extends benchmarking to interactive UI2Code, assessing VLMs’ ability to reproduce functional behaviors such as event handling and state transitions.

This manual dependency hinders large-scale random webpage generation and limits coverage of diverse interaction types.
Moreover, the benchmark’s evaluation paradigm—requiring explicit tagging of interactive elements and pre-specified commands—diverges from real user behaviors, emphasizing instruction following rather than true interaction reasoning. 
To overcome these issues, we propose UIFlow2Code, a scalable, flow-based benchmark that evaluates models’ ability to understand and reproduce multi-step interactions directly from visual and structural webpage states, eliminating the need for handcrafted task annotations and enabling behavior-grounded assessment.

\section{WebVIA Framework}

\begin{figure*}[t!]
    \centering
    \includegraphics[width=\linewidth]{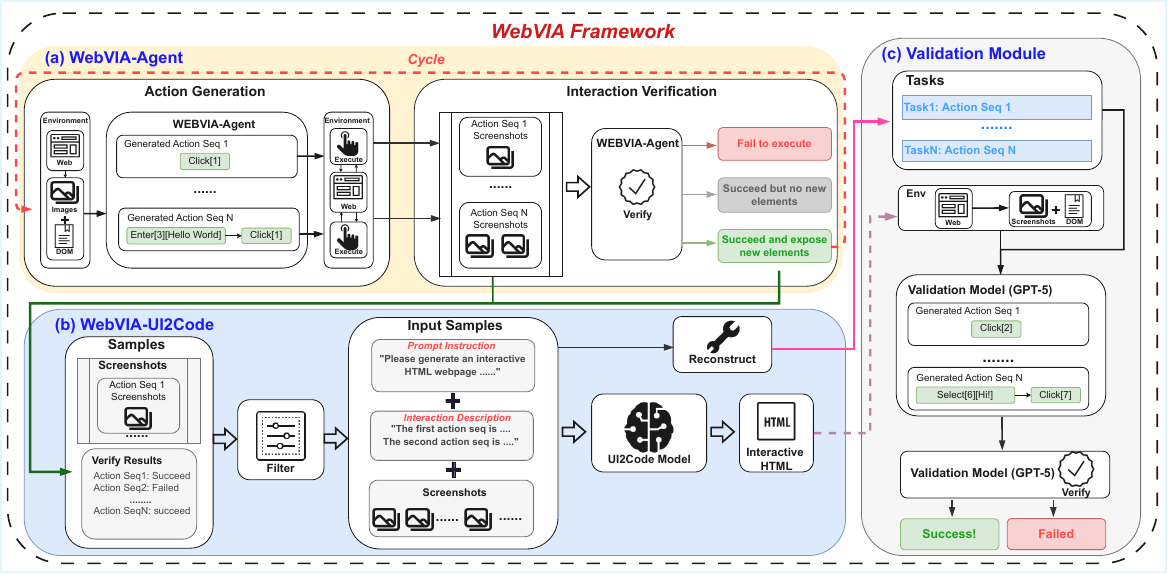}
    \vspace{-6mm}
    \caption{Overview of the \textbf{WebVIA} framework, which comprises three components: (a) an exploration agent to capture multi-state UI screenshots; (b) a UI2Code model to generate interactive code; (c) a validation module to verify the interactivity.}
    \label{fig:framework}
    \vspace{-5mm}
\end{figure*}

We propose WebVIA, the first agentic framework for interactive UI-to-Code generation and validation, designed to move beyond static rendering toward executable and verifiable front-end development. As illustrated in Figure~\ref{fig:framework}, WebVIA integrates three core components: an exploration agent that systematically interacts with HTML environments to uncover hidden states and produce validated UI screenshots, a UI2Code model that leverages these screenshots to generate executable code with both layout fidelity and interactivity, and a validation module that executes the generated code, verifies the support for intended GUI behaviors and functionalities. These components form a pipeline that bridges the gap between static UI rendering and robust, interactive front-end development.

\vpara{Problem Formulation.} We formalize interactive UI-to-Code generation as a sequential decision-making problem over a structured environment. The exploration agent interacts with a webpage environment $\mathcal{E}$. At each step $t$, the agent observes a multimodal state $s_t = (I_t, D_t)$, where $I_t$ is the rendered screenshot and $D_t$ is the DOM snapshot. The agent then selects and executes an action $a_t \in \mathcal{A}$ , and the environment transitions to a new state $s_{t+1}$. The action space $\mathcal{A}$ consists of standard web interactions, including clicks, text inputs, selections, and navigations. The agent progressively uncovers hidden states and constructs an interaction graph $\mathcal{G} = (\mathcal{S}, \mathcal{T})$, where $\mathcal{S}$ is the set of discovered states and $\mathcal{T}$ the set of verified transitions. 
Based on this interaction graph $\mathcal{G}$, the UI2Code model generates executable front-end code $\hat{C}$ that faithfully reproduces the interactive behaviors of the original environment. Finally, the validation module executes the generated code and assesses the interactivity of the rendered GUI by verifying whether it can still support the transitions $\mathcal{T}$ defined in the interaction graph $\mathcal{G}$.

\vpara{Environment.} The environment serves as the foundation of performing WebVIA framework. Inspired by prior work such as WebArena~\cite{zhou2023webarena}, we implement a dedicated web environment, denoted as \textit{WebEnv}, which renders a given HTML document within an isolated browser instance. The implementation builds on \texttt{GymAPI} for standardized interaction and \texttt{Playwright} for browser automation. \textit{WebEnv} supports three core capabilities: (1) rendering and capturing full-page screenshots, (2) extracting DOM trees with annotated interactive elements and the corresponding XPaths via JavaScript instrumentation, and (3) executing user interactions within the browser. These functionalities enable systematic grounding of both the visual and structural aspects of the webpage, providing a reliable basis for agent-driven exploration and evaluation.  

Real-world webpages are noisy and unstable due to advertisements, asynchronous loading, and external dependencies, making them unsuitable for controlled training and reproducible evaluation. To address this, we construct a large-scale synthetic environment in which webpages are automatically generated from templates and textual specifications. 
This design provides diversity and systematic coverage of common interaction patterns, ensures full controllability, and eliminates the unpredictability of real-world websites. The detailed synthesis pipeline is described in Appendix~\ref{app:env_synthesis}.

\subsection{Part 1:  Exploration Agent for Interactive UI Discovery}

The exploration agent is the core driver of WebVIA's interactive UI discovery process. It uncovers interactive elements within an HTML-based environment and constructs a reliable interaction graph of the user interface.  The agent has two key capabilities: \textbf{action generation} and \textbf{interaction verification}, and explores the GUI according to a perception–action–verification strategy: it grounds its understanding in rendered screenshots and DOM trees, proposes candidate interaction sequences, executes them in the environment, and verifies whether meaningful changes occur. This iterative process ensures comprehensive coverage of interactions and robustness against ineffective operations or redundant screenshots.

\vpara{Action Generation.} The exploration agent begins by proposing candidate interaction sequences that systematically cover potential user operations on a webpage. The agent, conditioned on its historical trajectory, observes the GUI state (i.e., the rendered screenshot and the DOM tree) and outputs action sequences. Each sequence can be a primitive operation (e.g., a single click) or a composite workflow (e.g., text entry followed by a button press). These sequences are executed within the environment to generate new candidate states for further exploration.

\vpara{Interaction Verification} To ensure efficiency and correctness, the agent verifies each executed action sequence by comparing the resulting state against the initial state according to the following criteria: (1) whether the action sequence successfully executed, and (2) whether new interactive elements appeared on the page. The comparison result falls into three categories: sequences that fail to execute (\textit{non-interactive}), sequences that succeed but reveal no new elements (\textit{usable} but not explored further), and sequences that succeed and expose new elements (\textit{usable} and retained for subsequent exploration). This interaction verification mechanism prunes redundant sequences and incorporates new interactive states into the exploration process.

\vpara{Exploration Strategy.} We propose a hybrid exploration strategy that integrates breadth-first and depth-first search within a perception–action–verification loop. At each iteration, the agent generates and executes candidate actions in parallel (breadth-first) to maximize coverage, while promising states are further expanded through depth-first exploration to uncover long-horizon workflows. As illustrated in Figure~\ref{fig:framework} (a), this process incrementally expands an interaction graph, where each validated state becomes the root of subsequent exploration. By balancing breadth and depth, WebVIA achieves both comprehensive coverage of interactive elements and efficient discovery of dynamic states, thereby constructing a reliable interaction graph for downstream code generation.

\subsection{Part 2: UI2Code Model for Interactive Front-End Code Generation} 

The ultimate objective of WebVIA is to move beyond static UI reconstruction and generate executable front-end code that faithfully captures the interactive functionality of the original interface. As illustrated in Figure~\ref{fig:framework}(b), we introduce a dedicated \textbf{UI2Code model} that is conditioned on the interaction graph derived from multiple screenshots produced by the exploration agent. Unlike prior approaches that rely on a single static screenshot, our model benefits from diverse interface states captured during exploration, enabling it to synthesize functionally coherent GUI components that support essential interactive behaviors.

\vpara{Multimodal Inputs.} Unlike prior approaches that rely on a single static screenshot, our UI2Code model is conditioned on multiple screenshots collected during exploration, along with their verified interaction relationships. This structured input captures both the visual layouts of diverse interface states and the causal transitions induced by user actions. Leveraging this enriched representation, the model learns to reason about dynamic workflows rather than treating the UI as a static rendering.

\vpara{Code Generation.} Conditioned on these multimodal inputs, the UI2Code model generates executable front-end HTML/CSS/JavaScript code. In contrast to prior approaches that only reconstruct static layouts, our model explicitly preserves interactive functionality, ensuring that the generated code is both visually faithful and behaviorally reliable.

\subsection{Part 3: Validation Module for Interactive Code}\label{sec:part3}
Existing evaluation methods for UI2Code models primarily focus on the visual fidelity of the generated interfaces, with little attention paid to verifying their interactivity. To address this limitation, the final component of WebVIA is the \textbf{validation module}, which ensures that the generated front-end code is functionally interactive. As shown in Figure~\ref{fig:framework}(c), we adopt a task-oriented validation procedure that directly evaluates the usability of the generated code. A set of tasks (e.g., filling out a form, submitting a query, or navigating to a target page) is pre-defined based on the interaction graph of the original GUI and executed on the synthesized interface. The synthesized interface passes validation if all tasks can be completed as intended. This evaluation goes beyond mere state--transition matching by providing a direct measure of whether the generated code supports coherent end-to-end user workflows.

\section{Training Methodology}

In this section, we describe how the two core models of WebVIA---WebVIA-Agent and WebVIA-UI2Code---are trained.

\subsection{WebVIA-Agent Training}

The exploration agent is trained on multimodal webpage states, consisting of rendered screenshots and filtered DOM trees. Its training objectives are threefold: (1) to ensure stability across diverse webpage layouts, (2) to comprehensively detect and interact with standard interactive elements, and (3) to jointly support the two core functions of action generation and interaction verification.

To support these objectives, we construct two complementary datasets. The \textit{Action Generation Dataset} contains pairs of webpage states and annotated interaction sequences. 
Each entry includes a screenshot, the corresponding DOM tree, and one or more ground-truth action sequences with their historical trajectories.The \textit{Interaction Verification Dataset} stores the screenshots before and after executing operation sequences, together with annotations indicating whether meaningful changes (such as new elements or layout updates) have occurred. Details of the dataset construction process are provided in Appendix~\ref{app:data_for_agent}.

The curated datasets are used to supervise the agent through \textit{supervised fine-tuning} (SFT) on \texttt{GLM-4.1V-9B-base}. The model learns to (i) predict valid action sequences from the action generation dataset, and (ii) classify meaningful transitions from the interaction verification dataset. Both tasks are optimized using cross-entropy loss. This joint training paradigm equips the agent with the ability to autonomously generate feasible interactions while reliably filtering out non-productive actions, providing a robust foundation for the WebVIA exploration pipeline.

\subsection{WebVIA-UI2Code Model Training}

The UI2Code model is trained to translate multiple UI screenshots into executable front-end code that preserves both layout fidelity and interactive functionality. Unlike conventional UI-to-code systems that rely solely on static screenshots, our training paradigm exploits the structured interaction graph collected by WebVIA, ensuring that the model learns to generate interactive HTML/CSS/JavaScript code.

We construct the \textit{WebView} dataset with 11k synthesized webpages, each paired with its ground-truth HTML/CSS/JavaScript code. For every webpage, the exploration agent systematically discovers states and transitions, producing an interaction graph $\mathcal{G}$ that contains rendered screenshots and validated action sequences. 
Instead of directly using the template-level HTML as supervision, we feed the exploration traces (multiple screenshots and their interaction graph) into Claude and generates the corresponding executable HTML/CSS/JavaScript code. The resulting (interaction graph, generated code) pairs form the core training data for our UI2Code model, ensuring that the supervision is aligned with the observed multimodal states and their verified interactions.

For fine-tuning, we adopt a structured prompt–response format, which is organized as: $\text{<think>}\cdots\text{</think><answer>}\cdots\text{</answer>}$. This formatting explicitly separates the reasoning context from the expected code output, enabling the model to learn stable mappings from multimodal observations to structured, executable code. The detailed data construction and data cases are provided in Appendix~\ref{app:data_for_ui2code}. We perform supervised fine-tuning (SFT) on \texttt{GLM-4.1V-9B-base} and \texttt{Qwen2.5-VL-7B-Instruct} on these formatted pairs.

\section{Experiments}

In this section, we conduct comprehensive experiments to validate the effectiveness of the proposed WebVIA framework. Our evaluation is organized around its two trainable components: the exploration agent and the UI2Code model. For the exploration agent, we examine both its intrinsic ability to generate and verify UI actions, as well as its performance in the full pipeline of interaction screenshot collection. For the UI2Code model, we evaluate its capability to generate interactive HTML/CSS/JavaScript code from multiple screenshots, focusing on structural fidelity and functional correctness.

\subsection{Evaluation Setup}
\vpara{Benchmark.} Since no public benchmark exists for evaluating our proposed WebVIA framework, we construct two dedicated benchmarks. \textbf{UIExplore-Bench} evaluates the exploration agent's ability to navigate complex webpages and collect interaction screenshots, while \textbf{UIFlow2Code-Bench} assesses the ability of UI2Code model to reconstruct webpages with both structural fidelity and functional correctness. All samples are carefully annotated to ensure accuracy and consistency. These benchmarks provide the first standardized protocol for this task and are designed to facilitate future research. Additional construction details of both UIExplore-Bench and UIFlow2Code-Bench are provided in Appendix~\ref{app:uiexplore-bench} and Appendix~\ref{app:uiflow2code-bench}, respectively.

\vpara{Baselines.} The WebVIA framework is designed to be model-agnostic, allowing both the exploration agent and the UI2Code model to be instantiated with any vision-language models. To systematically evaluate the advantages of our trained exploration agent and UI2Code model, we compare against a suite of state-of-the-art VLMs, including Claude-Sonnet-4, Claude-Sonnet-3.7, GPT-5, o4-mini, GPT-4o and Gemini-2.5-pro. These models can be seamlessly integrated into the \textsc{WebVIA} framework via API calls, without requiring any task-specific adaptation. The versions and API endpoints of baselines are provided in Appendix~\ref{app:baselines_version}.

\subsection{Single-Step Agent Evaluation} 
To assess the exploration agent’s performance independent of the full pipeline, we conduct single-step experiments on two fundamental tasks: \emph{action generation} and \emph{interaction verification}. 
The full prompt templates used for \textit{Action Generation} and \textit{Interaction Verification} are provided in Appendix~\ref{app:prompts}.

\vpara{Action generation.} In this task, the agent predicts a list of valid interactive elements from a given UI screenshot and DOM tree. 
We report \textit{Precision}, which measures the proportion of correctly predicted actions among the agent's selected actions, 
\textit{Recall}, which measures the proportion of correctly predicted actions with respect to the ground-truth actions, 
and \textit{F1}, which captures the harmonic mean of \textit{Precision} and \textit{Recall}.

As shown in Table~\ref{tab:precision_recall_f1_results}, the WebVIA-agent outperforms all baselines except Gemini-2.5-Pro. Its advantage stems from SFT training, which enables the agent to capture subtle structural patterns and focus on truly actionable elements. Different from Gemini-2.5-pro that tends toward an overly aggressive strategy of selecting nearly all visible elements, WebVIA-agent demonstrates a more balanced and reliable behavior. Furthermore, WebVIA-agent's predictions maintain high protocol fidelity, rarely producing mismatched DOM identifiers or invalid actions. By contrast, GPT-4o performs poorly, largely due to its inability to adhere to the prescribed interaction format, which undermines its applicability within the pipeline. The prompts

\begin{table}[h]
\centering
\small
\vspace{-2mm}
\caption{Comparison of action generation performance (Precision, Recall, and F1) on UIExplore-Bench with 87 action samples.}
\renewcommand{\arraystretch}{1.05}
\begin{tabular}{l|ccc}
\toprule
Model & Precision (\%) & Recall (\%) & F1 (\%) \\
\midrule
Gemini-2.5-pro                 & 74.01 & \underline{95.94} & 81.70 \\
GPT-5                          & \underline{81.66} & 88.41 & 81.85 \\
o4-mini                        & 79.01 & 91.80 & 83.16 \\
GPT-4o                         & 4.77  & 5.43  & 4.85 \\
Claude-Sonnet-3.7              & 75.29 & 95.18 & 81.72 \\
Claude-Sonnet-4                & 81.16 & 89.67 & \underline{83.38} \\
\midrule
WebVIA-Agent                   & \textbf{82.37} & \textbf{92.61} & \textbf{85.30} \\
\bottomrule
\end{tabular}
\vspace{-3mm}
\label{tab:precision_recall_f1_results}
\end{table}

\vpara{Interaction verification.} 
Given a set of images representing executed actions, the agent produces two Boolean outputs: \emph{"pass"}, indicating whether the sequence executes correctly, and \emph{"terminate"}, indicating whether the interaction introduces any new elements. We compute accuracy separately for both dimensions and use their average as the overall score. As shown in Table~\ref{tab:step_verification_results}, the WebVIA-agent achieves the best performance across all three metrics, demonstrating its superior verification ability. The superior terminate accuracy of WebVIA-agent underscores the effectiveness of SFT training in enhancing visual understanding and distinguishing genuinely new interactive elements. For example, when an interaction contains repeat elements of the same type (e.g., multiple “delete item” buttons), the agent successfully identifies the redundancy and terminates the exploration branch, thereby preventing unnecessary actions.

\begin{table}[h]
\centering
\small
\vspace{-2mm}
\caption{Comparison of verification performance across baseline VLMs and the WebVIA-agent on UIExplore-Bench with 53 verification samples.}
\setlength{\tabcolsep}{3.pt} % 调整列间距
\renewcommand{\arraystretch}{1.05}
\begin{tabular}{l|cc|c}
\toprule
Model & Pass Acc & Terminate Acc & Overall Acc \\
\midrule
Gemini-2.5-pro         & 94.34 & 81.13 & 87.74 \\
GPT-5                  & \underline{96.23} & \underline{84.91}& \underline{90.57} \\
o4-mini                & 94.34 & 83.02 & 88.68 \\
GPT-4o                 & 33.96 & 62.26 & 48.11 \\
Claude-Sonnet-3.7      & 94.34 & 77.36 & 85.85 \\
Claude-Sonnet-4        & 94.34 & 77.36 & 85.85 \\
\midrule
WebVIA-Agent           & \textbf{98.11} & \textbf{86.79} & \textbf{91.51} \\
\bottomrule
\end{tabular}
\vspace{-4mm}
\label{tab:step_verification_results}
\end{table}

\subsection{Pipeline-Level Agent Evaluation}

While single-step experiments isolate the agent's capabilities on action generation and interaction verification, they fail to provide a comprehensive assessment of its effectiveness in realistic, end-to-end scenarios. We further evaluate the exploration agent within the WebVIA framework, where it autonomously explores webpages, generates interaction traces, and collects representative screenshots.

\vpara{Evaluation metrics.} To assess the effectiveness of the exploration agent throughout the entire collection of interaction screenshots, we adopt three complementary metrics: \textbf{Completeness} measures the coverage of action generation, i.e., the proportion of distinct UI elements successfully explored. \textbf{Correctness} quantifies the correctness of verification results, reflecting the agent’s ability to determine whether an executed action achieves its intended effect. \textbf{Deduplication Rate} quantifies the prevalence of redundant or repeated actions within the generated traces, serving as an indicator of exploration efficiency. We compute an overall score as a weighted combination of the three metrics:  
\begin{equation}
\small
\text{Overall} = 0.40 \cdot \text{Comp}
+ 0.35 \cdot \text{Correct}
+ 0.25 \cdot \text{Dedup}
\label{eq:overall}
\end{equation}
, where the weights are empirically determined to balance coverage, correctness, and efficiency.

\vpara{Evaluation results.} As reported in Table~\ref{tab:agent_results}, WebVIA-Agent achieves the best overall score of 89.8\%, surpassing all baseline models. It achieves the best Completeness (93.1\%) and Correctness (97.7\%), confirming its ability to both discover diverse actionable elements and reliably validate their outcomes. These improvements arise from supervised fine-tuning, which strengthens structural understanding and encourages a balanced exploration strategy. 
For instance, when encountering redundant screenshots with overlapping content, WebVIA-Agent prioritizes unexplored regions and reducing duplication. By contrast, models such as o4-mini and Gemini-2.5-pro often re-trigger identical actions (e.g., repeatedly clicking the same button), yielding inefficiencies despite high nominal coverage.

\begin{table*}[t!]
\centering
\small
\vspace{-2mm}
\caption{Pipeline-level performance comparison across baseline VLMs and the proposed WebVIA-Agent on UIExplore-Bench with 56 webpages.}
\renewcommand{\arraystretch}{1.05}
\begin{tabular}{l|cccc}
\toprule
Model & Completeness (\%) & Correctness(\%) & Deduplication Rate (\%) & Overall Score (\%) \\
\midrule
Gemini-2.5-pro        & \underline{92.61} & \underline{95.39} & 5.60  & 71.83 \\
GPT-5                & 76.66 & 90.19 & \underline{93.82} & 85.69 \\
o4-mini               & 91.73 & 94.07 & 52.73 & 82.80 \\
GPT-4o                & 16.46 & 62.63 & 97.45 & 52.87 \\
Claude-Sonnet-3.7   & 75.86 & 94.06 & 72.36 & 81.35 \\
Claude-Sonnet-4     & 86.26 & 95.07 & 80.36 & \underline{87.87} \\
\midrule
WebVIA-Agent               & \textbf{93.12} & \textbf{97.71} & \textbf{72.73} &\textbf{89.63} \\
\bottomrule
\end{tabular}
\vspace{-3mm}
\label{tab:agent_results}
\end{table*}

\subsection{Interactive Code Generation Evaluation}
To assess the capability of UI2Code model in generating fully interactive HTML/CSS/JavaScript code, we extend the evaluation beyond conventional static UI2Code task to interactive code generation. 

\vpara{Evaluation metrics.} For each generated HTML page, a set of \textit{tasks} is defined based on the corresponding input images. 
Each \textit{task} undergoes the validation process and is labeled as either \textit{pass} or \textit{fail}. 
The final evaluation metric is calculated as the ratio of the number of \textit{pass} tasks to the total number of \textit{tasks}.

\vpara{Evaluation results.} As reported in Table~\ref{tab:UI2Interact}, supervised fine-tuning on the \texttt{WebView} dataset substantially improves the ability of both Qwen-2.5-VL-7B and GLM-4.1V-9B to generate executable interactive HTML/CSS/JavaScript code, whereas their base counterparts fail to produce valid outputs. This contrast highlights that interactive training data are indispensable for enabling interaction capabilities. Interestingly, although our supervised fine-tuning is conducted exclusively on interactive UI2Code data, we also observe consistent improvements on static UI2Code benchmarks. This suggests that interactive training data provide richer structural and functional supervision than conventional single-state screenshots, thereby enhancing the model’s ability to capture layout fidelity and semantic alignment even in static scenarios. Details of the evaluation prompt for interactive code generation are provided in Appendix~\ref{app:ui2code_prompts}.

\begin{table}[h]
\small
\vspace{-2mm}
\centering
\caption{Performance comparison on static (Design2Code) and interactive UI2Code (UIFlow2Code).}
\renewcommand{\arraystretch}{1.05} % 调大行距
\setlength{\tabcolsep}{3.pt} % 调整列间距
\begin{tabular}{c|c|c}
\toprule
Model  & Design2Code    & UIFlow2Code \\
\midrule
Gemini-2.5-pro          &  89.5          & \underline{90.2} \\
GPT-5                   &  \underline{89.7}          & 69.9 \\
o4-mini                 &  63.8          & 69.4 \\
GPT-4o                  &  35.3          & 59.3\\
Claude-Sonnet-3.7       &  77.7          & 81.5 \\
Claude-Sonnet-4         &  81.2          & 82.1 \\
\midrule
Qwen2.5-VL-7B-Instruct  & 29.1    & / \\
WebVIA-UI2Code-Qwen                & \textbf{34.3}     & \textbf{75.9} \\
\midrule
GLM-4.1-V-9B-Base       & 58.3    & / \\
WebVIA-UI2Code-GLM                & \textbf{63.0}    & \textbf{84.9} \\
\bottomrule
\end{tabular}
\vspace{-1mm}
\label{tab:UI2Interact}
\end{table}

\subsection{Average Trace Length of WebVIA-Agent}

\begin{figure}
    \centering
    \includegraphics[width=0.95\linewidth]{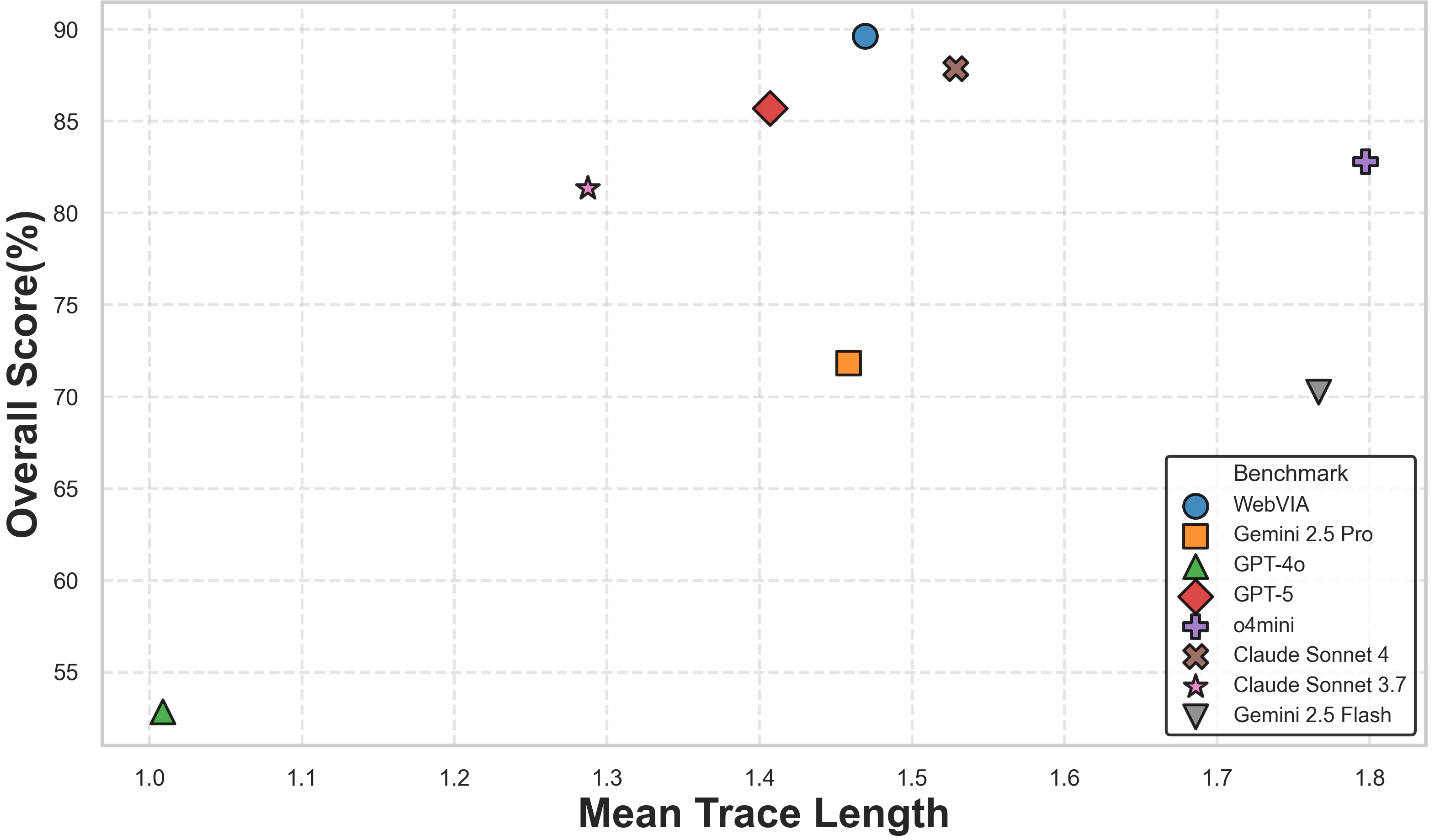}
    \caption{Correlation between the mean interaction trace length and the overall exploration score across our WebVIA-Agent and various VLMs.}
    \vspace{-3mm}
    \label{fig:trace}
\end{figure}

In our ablation study, we analyze the relationship between average trace length and overall performance as reported in Table~\ref{tab:agent_results}. Here, trace length refers to the number of interaction steps executed by the pipeline-level WebVIA-Agent during a full webpage exploration. A lower trace length may indicate more intelligent action planning, but it may also reflect premature termination that fails to capture deeper interactive elements. As shown in Figure~\ref{fig:trace}, WebVIA-Agent achieves a balanced mean trace length together with the highest overall performance. This combination indicates WebVIA-Agent is not only efficient but also consistently effective, indicating that it strikes a strong balance between quality and speed.

\section{Conclusion}

In this work, we present WebVIA, an agentic framework for interactive UI-to-code generation and validation.
Unlike prior methods limited to static HTML/CSS reconstruction, WebVIA introduces an \textit{exploration–generation–validation} pipeline that enables interaction-aware, behavior-preserving code synthesis.
Built upon large-scale GUI interaction and WebView data, two specialized agents—an exploration agent and a UI2Code generator—jointly produce executable and verifiable web interfaces.

\section*{Limitations}
Although WebVIA establishes a new paradigm for interactive UI-to-Code generation, there remain two limitations in scalability and generalization that need to be addressed before achieving broader applicability. (1) In the exploration stage of WebVIA pipeline, the action types are restricted to Click, Enter, and Select. Executing broader action types such as Drag and Draw requires precise pixel coordinates, which defers from our current approach of ID based DOM to XPath execution. (2) Training the agent primarily on synthetic Webpages may limit its ability to handle certain specialized interaction tasks in real-world settings. For example, WebVIA-Agent struggles with domains such as calculators or function-plotting interfaces, where interaction patterns deviate substantially from the structures observed in the training environment. These constraints delineate the current scope of WebVIA and point to concrete directions for extending its applicability in future research.


\bibliography{custom}

\clearpage

\appendix

\section{Appendix}
\label{sec:appendix}

\subsection{Environment HTML Synthesis}\label{app:env_synthesis}

\begin{figure}[h]
    \centering
    \includegraphics[width=1\linewidth]{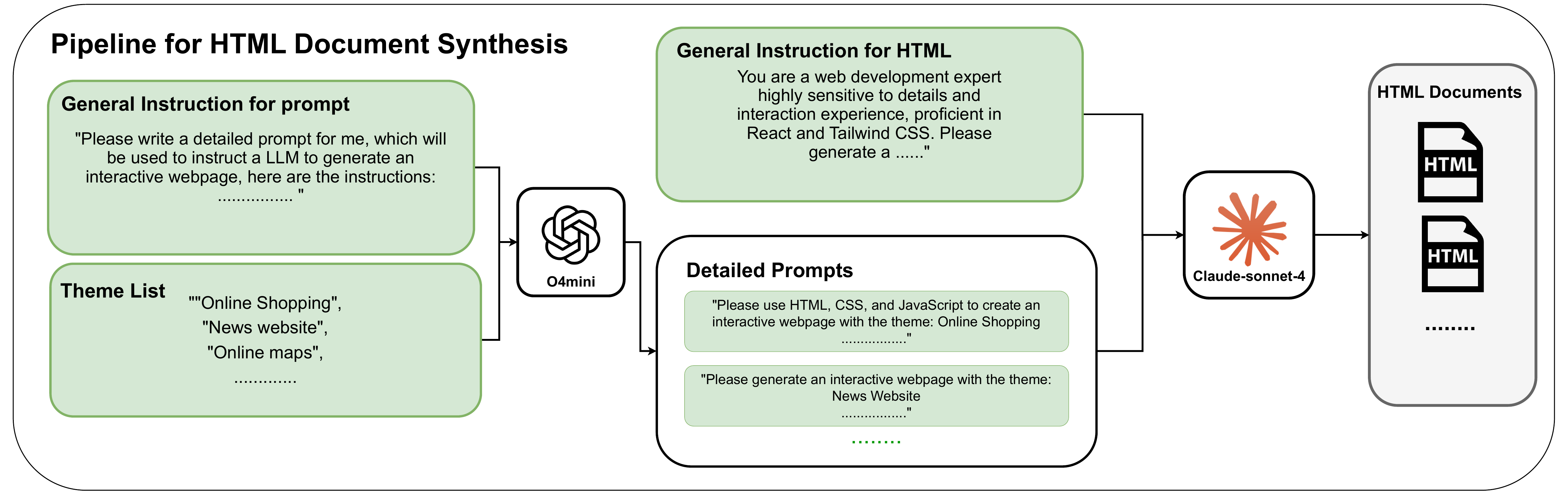}
    \caption{Overview of the webpage synthesis process in the WebVIA framework. }
    \label{fig:env_synthesis}
\end{figure}

To enable scalable data generation for agent training, we construct a simulated HTML synthesis pipeline (See Figure \ref{fig:env_synthesis}) that automatically produces diverse and interactive webpage environments.
The pipeline begins with a theme list (e.g., online shopping, news websites, online maps), from which a general instruction template is provided to generate task-specific prompts.
Using o4-mini, we expand each general instruction into detailed natural-language prompts that specify the structure, layout, and interaction requirements for webpages.
These detailed prompts are then fed into Claude-Sonnet-4, which generates executable HTML/CSS/JavaScript documents formulated within the React framework (hereafter referred to as HTML documents/HTML codes), enabling the construction of interactive elements such as buttons, input forms, and navigation menus.
% 可以改成React框架下的html code
This two-stage generation process ensures semantic diversity (via high-level theme variation) and functional richness (via prompt-guided interaction synthesis).
The resulting webpages form a large-scale synthetic environment for the WebVIA-Agent, supporting consistent and reproducible training across varied interface types and behaviors.

\vpara{Webpage Design Instruction Template.} To guide the generation of diverse interactive webpages, we design a general instruction template that can be automatically adapted to different themes in the synthesis pipeline.
Each theme (e.g., Online Shopping, News Website, Online Maps, Portfolio Page) is inserted into the template to form a specific prompt.
The following template illustrates the general structure used to generate detailed webpage instructions, as shown in Figure~\ref{fig:webpage-design-template}.

\begin{figure}[t]
\centering
\begin{tcolorbox}[
    colback=gray!2,
    colframe=blue!70!black,
    colbacktitle=blue!12!white,
    title=\textbf{Webpage Design Instruction Template},
    fonttitle=\bfseries,
    coltitle=blue!50!black,
    enhanced,
    sharp corners,
    boxrule=0.5pt,
    left=5pt,
    right=5pt,
    top=5pt,
    bottom=5pt,
    titlerule=0mm,
    width=0.5\textwidth,
    title style={left color=blue!8!white, right color=blue!5!white},
]
\small
Please write a detailed prompt that will be used to instruct a text-to-image model to generate an interactive webpage HTML code with the theme \textbf{``<INSERT THEME FROM THEME LIST>''}.

\vspace{2mm}

\textbf{Specific Requirements:}
\begin{enumerate}
    \item The webpage content should revolve around the specified theme and include a wide variety of theme-related modules.
    \item The webpage must contain multiple interactive elements, limited to \texttt{buttons}, \texttt{input fields}, and \texttt{dropdown selectors}. Each interactive component should cause corresponding and reasonable changes on the webpage.
    \item The webpage content should be rich, detailed, and contextually diverse.
    \item The output should only contain the final prompt for the AI to generate the webpage—without explanations, metadata, or additional commentary.
\end{enumerate}
\end{tcolorbox}
\caption{Template used to construct webpage design prompts for generating interactive webpage HTML code.}
\label{fig:webpage-design-template}
\end{figure}

\vpara{Code Generation Prompt.}
To ensure functional completeness and visual consistency, we design a dedicated code generation prompt that explicitly instructs the model to generate self-contained and interaction-ready HTML code.
Each generated page is automatically executed and rendered in a browser environment using \textit{Playwright} to verify both visual correctness and interactive functionality. 
Only webpages that successfully render and execute without errors are retained, ensuring that the synthesized dataset is composed of valid, executable, and behaviorally rich webpages. 
The following prompt defines the instruction used for generating executable HTML code, as illustrated in Figure~\ref{fig:code_generate_prompt}.

\begin{figure*}[t!]
\centering
\begin{tcolorbox}[
    colback=gray!2,
    colframe=blue!70!black,
    colbacktitle=blue!12!white,
    title=\textbf{Code Generation Prompt},
    fonttitle=\bfseries,
    coltitle=blue!50!black,
    enhanced,
    sharp corners,
    boxrule=0.5pt,
    left=6pt,
    right=6pt,
    top=5pt,
    bottom=5pt,
    titlerule=0mm,
    width=0.96\textwidth, % ✅ 跨双栏
    title style={left color=blue!8!white, right color=blue!5!white}
]
\small
You are a web development expert highly sensitive to details and interaction experience, proficient in React and Tailwind CSS. 
Please generate a highly interactive single-page application with reasonable layout and rich content for the specified theme according to the following requirements.

\vspace{2mm}
\textbf{Basic Requirements:}
\begin{enumerate}
    \item Generate a complete interactive single-page website rendered using \textbf{React (v18)} and \textbf{Tailwind CSS (v3+)}. 
    \item Return only the full source code wrapped within \texttt{<html>...</html>} tags. \textbf{Do not} include markdown wrappers, explanations, or code comments.
    \item Must include the following dependencies:
\begin{lstlisting}[basicstyle=\ttfamily\footnotesize, frame=none, breaklines=true]
<script src="https://cdn.jsdelivr.net/npm/react@18.0.0/umd/react.development.js"></script>
<script src="https://cdn.jsdelivr.net/npm/react-dom@18.0.0/umd/react-dom.development.js"></script>
<script src="https://cdn.jsdelivr.net/npm/@babel/standalone/babel.js"></script>
<script src="https://cdn.tailwindcss.com"></script>
<link rel="stylesheet" href="https://cdnjs.cloudflare.com/ajax/libs/font-awesome/5.15.3/css/all.min.css"></link>
\end{lstlisting}
\end{enumerate}

\vspace{-1mm}
\textbf{Interactivity and Functional Areas:}
\begin{enumerate}
    \item All interactive components (\texttt{input}, \texttt{button}, \texttt{select}) must trigger meaningful updates to the rendered page.
    \item For editable content, use modals, dropdowns, or input forms with complete validation.
    \item Use real pictures from \url{https://picsum.photos/}. Each image must have a fixed URL and remain constant across reloads.
\end{enumerate}

\vspace{-1mm}
\textbf{Page Structure and Layout:}
\begin{enumerate}
    \item Include logical partitions (navigation, sidebar, main content, etc.) referencing modern app layouts.
    \item Ensure all sections are populated; empty placeholders are not allowed.
    \item The visual style must match the assigned theme (e.g., business, minimalism, tech, lifestyle).
\end{enumerate}

\vspace{-1mm}
\textbf{Notes:}
\begin{itemize}
    \item Do not output explanations or text outside the code.
    \item Ensure all theme-related UI logic is complete and intuitive.
\end{itemize}

\vspace{1mm}
\textbf{Webpage Description:} \texttt{<INSERT DETAILED PROMPT FROM STAGE 1>}
\end{tcolorbox}
\caption{Code Generation Prompt used for large-scale HTML synthesis.}
\label{fig:code_generate_prompt}
\end{figure*}

% code generation prompt by LLM

\subsection{Training Dataset for Exploration Agent}\label{app:data_for_agent}

To train the WebVIA-Agent for robust and generalizable UI exploration, we construct a large-scale GUI interaction dataset derived from the synthetic HTML environments described in Section~\ref{app:env_synthesis}. 
Each webpage instance provides a structured environment where the agent can perceive the DOM tree, rendered screenshot, and interaction history, enabling the model to learn both visual and semantic representations of interactive elements.

\vpara{Automated Data Construction.}
To efficiently construct the training data for the WebVIA-Agent, we design an automated data generation pipeline that produces two complementary datasets: \textbf{(1) Action Generation} and \textbf{(2) Interaction Verification}. 
Both are generated using the \textit{o4-mini} model within the \textit{WebEnv} environment, which supports both synthetic and real webpages.

For the \textbf{Action Generation} dataset, \textit{o4-mini} serves as a general-agent and is executed once across the entire WebVIA environment, with its exploration trajectories recorded and subsequently reconstructed. The reconstruction format preserves both the historical context and the input (a paired visual--structural state, consisting of the rendered UI screenshot and its associated DOM hierarchy), while the output is represented as sequences of operations (e.g., \texttt{boxed\{click[1]\}}, \texttt{boxed\{enter[2][Hello World!], click[5]\}}).

For the \textbf{Interaction Verification} dataset, the pipeline executes these action sequences. Each sequence $a_{t:t+k}$ may consist of multiple actions, where each action produces an intermediate state $s_{t+1}, s_{t+2}, \ldots, s_{t+k}$. The resulting set of screenshots across these successive states is jointly considered as the post-action evidence. Formally, each interaction tuple is represented as
\[
(s_t, a_{t:t+k}, \{s_{t+1}, s_{t+2}, \ldots, s_{t+k}\}, r_t),
\]
where $r_t \in \{0,1\}$ indicates the correctness of the overall outcome.  
This dual-branch data generation process yields approximately \textbf{180K} verified interaction samples across \textbf{20K} webpages, encompassing a wide range of UI components, event bindings, and layout hierarchies.

% Verification: 在我们WebVIA 中跑了一整遍，然后把Verfication保存并重构，重构的格式 包含输入（UI screenshot + DOM） 输出(一组操作序列，比如boxed {click[1]}, boxed {enter[2][Hell0 World!], click[5]} ) 

%Human in the loop verification 这里 “annotators review the visual and DOM differences between” 对于verification task，输入只有图片没有dom difference，删除Dom differences
\vpara{Human-in-the-Loop Verification.}
To ensure data reliability, we incorporate a semi-automated verification stage, where annotators rely on rule-based checks to assess both the correctness of generated action sequences and the corresponding success labels. 
For action sequences, automated filtering is applied to remove cases with overly short selections or high redundancy, while inconsistent examples are corrected when necessary. 
Verification is handled through a mixed procedure: annotators first sample and manually inspect a subset of instances, and their findings are used to identify recurring types of outcomes that diverge from human judgment. These patterns are then formalized into rules, which guide the selective removal or adjustment of the affected cases.

\vpara{Discussion.} The resulting dataset unifies both \textit{action generation} and \textit{interaction verification} supervision, encouraging WebVIA-Agent to reason not only about what actions to take but also about their functional outcomes.
Despite being primarily trained on synthetic webpages, the agent exhibits strong generalization to real-world sites, successfully handling unseen layouts, interaction patterns, and DOM structures.

% 这里写一下 我们数据怎么得到的以及action generation data compare data cases (数据形式)

% 这里可以写一下我们的训练数据是怎么组织的

\subsection{Training Dataset for Interactive UI2Code Model}\label{app:data_for_ui2code}

% 写一下 我们合成数据的流程 可以画个figure （这里可以简单画 和你之前ppt差不多就行）

% 写一下数据cases 给几个cases

To enable the WebVIA-UI2Code model to generate executable and interactive HTML code,
we construct the \textit{WebView} dataset, which aligns multi-state UI screenshots with their corresponding ground-truth interactive webpages.
Each data instance captures both the static visual layout and the dynamic behavioral transitions of a webpage, providing comprehensive supervision for learning interaction-aware code generation.

\begin{figure}[t!]
    \centering
    \includegraphics[width=1\linewidth]{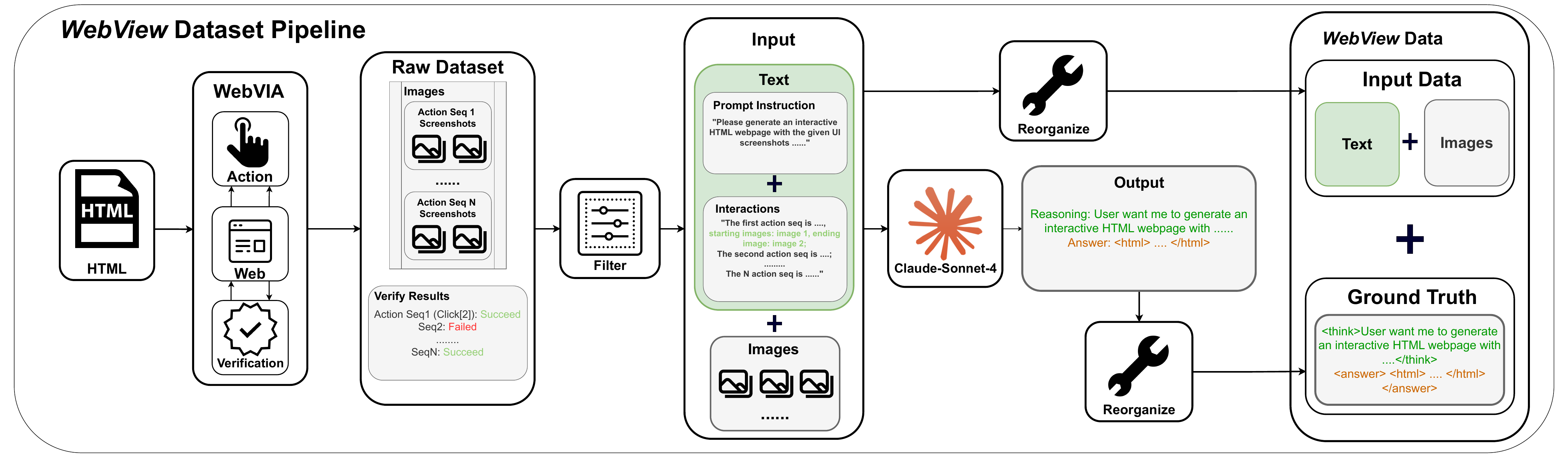}
    \caption{Pipeline for constructing the \textit{WebView} dataset. }
    \label{fig:ui2code_data_pipeline}
\end{figure}

\vpara{Data Generation Pipeline.} 
As illustrated in Figure~\ref{fig:ui2code_data_pipeline}, 
the construction of the \textit{WebView} dataset follows a three-stage pipeline: (1) \textbf{Webpage Construction.} Using the HTML synthesis pipeline described in Appendix~\ref{app:env_synthesis}, we generate a large number of interactive webpages with diverse themes (e.g., shopping, news, portfolio, dashboard). Each page contains multiple interactive elements such as buttons, input fields, dropdowns, and forms, all bound with interactive behaviors. (2) \textbf{State Exploration.} The WebVIA-Agent interacts with each synthesized page to traverse all reachable states. During this process, it captures multi-state screenshots $\{I_1, I_2, ..., I_n\}$ together with corresponding DOM snapshots and event logs, reflecting visual and structural transitions triggered by user interactions. (3)  \textbf{Interactive Code Generation.} After obtaining multi-state UI screenshots and corresponding interaction traces from the exploration stage, we adopt a multimodal instruction–response formulation to transform these visual observations into executable code.
Specifically, we prompt the Claude-Sonnet-4 model to generate code under a structured reasoning format that explicitly separates the thought process and the final output using the tags \texttt{<think>} and \texttt{<answer>}.
Within the \texttt{<think>} block, the model is encouraged to analyze the provided screenshots and interaction logs, infer component hierarchies, and reason about event dependencies.
The final HTML implementation is then produced in the \texttt{<answer>} block, ensuring a clear delineation between reasoning and generation.
This design allows the model to perform interpretable, step-by-step reasoning about webpage structure and interactivity before emitting executable code, leading to more functionally correct, visually coherent, and behavior-consistent outputs.

\vpara{Prompt for Interactive Code Generation.}
To ensure consistent reasoning and interpretable generation during interactive code synthesis, we design a multimodal instruction prompt tailored for \textsc{Claude-Sonnet-4}, as illustrated in Figure~\ref{fig:interactive_code_generate_prompt}.

\begin{figure*}[t!]
\centering
\begin{tcolorbox}[
    colback=gray!2,
    colframe=blue!70!black,
    colbacktitle=blue!12!white,
    title=\textbf{Interactive Code Generation Prompt},
    fonttitle=\bfseries,
    coltitle=blue!50!black,
    enhanced,
    sharp corners,
    boxrule=0.5pt,
    left=5pt,
    right=5pt,
    top=5pt,
    bottom=5pt,
    titlerule=0mm,
    title style={left color=blue!8!white, right color=blue!5!white},
    width=\textwidth
]
\small
You are highly skilled at building interactive webpages with React and Tailwind, and can precisely reconstruct a complete HTML interactive webpage based on multiple webpage screenshots provided by the user.
\vspace{2mm}

\textbf{Initial Interface Requirements:}
\begin{enumerate}
    \item Build the page strictly according to the first webpage screenshot provided by the user. It must be exactly the same as the first screenshot you receive.
    \item Do not miss any details. Background colors, fonts, font sizes, spacing, borders, icons, and text must strictly match the screenshot.
    \item Every line of text in the screenshot must be presented verbatim.
    \item For images, please use real pictures from the \url{https://picsum.photos/} library, with URLs like \url{https://picsum.photos/id/.../.../...}. \textbf{Each image must explicitly list its URL.} Do not use reusable image components. Each webpage component’s image URL must be fixed and must not be randomly regenerated each time.
\end{enumerate}

\textbf{Task Requirements:}
\begin{enumerate}
    \item The user will send you multiple images. Each image represents a screenshot of the webpage after a single interactive operation, and all images together represent the screenshots resulting from all operations performed on this page.
    \item The user will send you a detailed operation-sequence list. Each item in the list represents one operation sequence and will tell you which image (by index in the images you received) is the starting image (the page before the operation), and which images correspond to the sequence of screenshots after each step in the operation. Locate these images yourself. An operation sequence may include multiple operations, i.e., it may span multiple images. Some operation sequences have many intermediate steps, but among the images only the first and the last are provided—identify the specific operation content yourself.
    \item After locating these images, read the operation description for that item. There are three types of operations: “input,” “click,” and “select.” Correctly identify which interactive component in the screenshots corresponds to each operation, and implement them correctly in the generated HTML webpage.
    \item All interactive operations given to you must be perfectly replicated in the generated HTML, meaning they must be fully functional, and once completed, the page must match the corresponding screenshots.
\end{enumerate}

\textbf{Library Requirements:}
\begin{lstlisting}[basicstyle=\ttfamily\footnotesize, breaklines=true, frame=none]
<script src="https://cdn.jsdelivr.net/npm/react@18.0.0/umd/react.development.js"></script>
<script src="https://cdn.jsdelivr.net/npm/react-dom@18.0.0/umd/react-dom.development.js"></script>
<script src="https://cdn.jsdelivr.net/npm/@babel/standalone/babel.js"></script>
<script src="https://cdn.tailwindcss.com"></script>
<link rel="stylesheet" href="https://cdnjs.cloudflare.com/ajax/libs/font-awesome/5.15.3/css/all.min.css"></link>
\end{lstlisting}
You may use Google Fonts.

\vspace{2mm}

\textbf{Code Output Format:}
\begin{enumerate}
    \item Output only the code within the complete \texttt{<html></html>} tags.
    \item Do not add markdown quotes or “html” before or after the code.
\end{enumerate}
\end{tcolorbox}
\caption{The instruction prompt used for interactive code generation within the \textsc{WebVIA} framework.}
\label{fig:interactive_code_generate_prompt}
\end{figure*}

\vpara{Quality Assessment.} 
To ensure that the synthesized interactive code is functionally executable and visually coherent, 
each generated HTML file is automatically rendered in a browser environment powered by \textit{Playwright}.
The automatic check focuses on whether the page can be successfully loaded and rendered without runtime errors. 
For interactive components and state transitions, we rely on sample-based human verification: annotators review a subset of interactions (e.g., clicking, text input, and selection), and their inspection confirms that the vast majority of components behave as expected. 
This combined strategy allows us to filter out pages that fail to render while providing evidence that the generated interactions remain largely reliable, ensuring high-quality supervision for training the WebVIA-UI2Code model.

\subsection{UIExplore-Bench Construction}\label{app:uiexplore-bench}

To systematically evaluate the performance of the WebVIA-Agent, we construct \textit{UIExplore-Bench}, a benchmark specifically designed for assessing both fine-grained interaction reasoning and end-to-end exploration performance. 
Unlike previous web agent datasets that primarily target task completion, UIExplore-Bench focuses on measuring the agent’s capability to recognize UI components, execute valid interactions, and verify their functional correctness.

\vpara{Benchmark Composition.}
UIExplore-Bench comprises three complementary subsets:
\begin{itemize}
    \item \textbf{Action Generation Set.} This subset contains \textbf{87} annotated interaction samples, each represented as paired \textit{(image, DOM tree)} inputs. 
    Each sample specifies a target action in natural language (e.g., “click the \texttt{Search} button” or “enter text in the input field”), enabling quantitative evaluation of the agent’s action prediction accuracy given a specific webpage state.
    
    \item \textbf{Interaction Verification Set.} This subset includes \textbf{53} verification cases, each consisting of pre- and post-interaction screenshots along with DOM snapshots. 
    The task requires the agent to determine whether the executed action produces a functionally valid change (e.g., modal opening, content update, navigation), thereby assessing its ability to reason about dynamic webpage transitions.

    \item \textbf{Pipeline-Level Evaluation Set.} We build a larger-scale evaluation suite containing \textbf{56} complete webpages.
    For each webpage, the agent must autonomously explore all interactive components, generate valid action traces, and collect representative screenshots across multiple UI states.
    This subset evaluates the agent’s full exploration pipeline—from perception and action generation to interaction validation.
\end{itemize}

\vpara{Construction Pipeline.} UIExplore-Bench comprises three subsets: an \textit{Action Generation Set}, an \textit{Interaction Verification Set}, and a \textit{Pipeline Evaluation Set}.
Each subset targets a distinct dimension of GUI reasoning and follows a dedicated data construction pipeline to ensure reliability and coverage.
\begin{itemize}
    \item \textbf{Action Generation Set.} This subset focuses on evaluating the agent’s ability to generate valid interaction actions from a single webpage state. We first select 100 webpages across diverse themes and employ the WebVIA-Agent to collect all possible interaction screenshots within each page. For every UI $<screenshot, DOM>$ pair, the leading models such as Claude-Sonnet-4, GPT-5, and Gemini-2.5-Pro 
    +re are prompted to generate candidate interaction sequences. From each webpage, we retain the UI screenshot associated with the richest interaction set and aggregate all model-generated actions. Subsequently, human annotators manually verify the combined sequences to remove non-functional, ambiguous, or nonexistent interactions, resulting in a high-quality ground-truth action set.

    \item \textbf{Interaction Verification Set.} This subset targets the evaluation of an agent’s ability to verify whether an interaction has been successfully executed. We reselect 100 webpages spanning diverse application domains and employ the WebVIA-Agent solely to collect paired pre- and post-interaction states, including screenshots and action logs. Subsequently, human annotators manually inspect and label these pairs to determine whether the executed interaction leads to a valid and functionally consistent state transition. After filtering out ambiguous or redundant cases, 53 high-quality samples are retained as ground-truth verification data, each providing a reliable reference for assessing interaction correctness.

    \item \textbf{Pipeline Evaluation Set.} This subset is designed to evaluate the agent’s end-to-end exploration capability within complete webpage environments. We re-select 100 webpages across diverse domains and conduct autonomous exploration using the leading models such as Claude-Sonnet-4, GPT-5, and Gemini-2.5-Pro within the WebVIA framework. For each webpage, we merge all screenshots from different models and perform manual annotation to filter out invalid, overly long, or non-existent interaction elements. After this refinement process, 56 webpages are retained as high-quality ground-truth cases.

\end{itemize}

% We construct UIExplore-Bench through a semi-automatic pipeline combining model-assisted generation and human verification.
% Initially, a large pool of interactive webpages is synthesized using the HTML generation pipeline described in Appendix~\ref{app:env_synthesis}. 
% The WebVIA-Agent then performs controlled exploration within each environment to produce candidate samples containing screenshots, DOM trees, and interaction logs.
% Next, the \textit{o4-mini} model generates natural-language action annotations and success labels based on DOM diffs and visual changes.
% Finally, human annotators manually verify ambiguous or failure cases to ensure semantic correctness and consistency across samples.

% 这里详细写一下uiexplore-bench constuction 

\subsection{UIFlow2Code-Bench Construction}\label{app:uiflow2code-bench}

To systematically evaluate interactive UI-to-Code generation, we construct \textit{UIFlow2Code-Bench}, 
a benchmark designed to assess a model’s ability to generate executable, behavior-preserving HTML code 
from multi-state user interface (UI) observations. 
Unlike existing UI2Code benchmarks such as Design2Code~\cite{si2024design2Code} and FullFront~\cite{sun2025fullfront}, 
which focus solely on static layout reconstruction, 
UIFlow2Code-Bench explicitly incorporates state transitions and interaction traces, 
enabling fine-grained evaluation of interaction-aware code synthesis.

\vpara{Benchmark Composition.} UIFlow2Code-Bench contains \textbf{50} synthesized webpages covering diverse domains including e-commerce, news, dashboards, and portfolio sites. 
Each sample is composed of (1) a sequence of multi-state UI screenshots captured during interaction, 
and (2) the corresponding executable ground-truth HTML implementation.
On average, each webpage includes 4–6 interaction states and 8–12 functional components such as buttons, modals, forms, and dropdown menus. 
These paired multi-view samples enable fine-grained evaluation of interaction-aware code synthesis.

\vpara{Construction Pipeline.} We select 100 webpages across diverse domains and employ the WebVIA framework to conduct systematic exploration using the WebVIA-Agent. 
For each webpage, the agent traverses all available interactive components and records corresponding multi-state UI screenshots. 
Human annotators then manually select 6 representative screenshots per webpage, each associated with 2 to 5 interaction actions, 
covering diverse visual layouts and interaction task types such as clicking, text input, and selection. 
These selected UI states collectively define the interaction trajectories that the model is expected to reproduce. 
In the evaluation phase, the generated HTML code is considered correct if it can faithfully execute the annotated actions and reproduce the corresponding state transitions, 
rather than merely replicating static visual appearance. 
This design ensures that UIFlow2Code-Bench emphasizes \textit{interaction consistency} and functional correctness over superficial layout matching.

% 这里详细写一下uiflow2code-bench constuction 

\subsection{Baselines Versions and API Endpoints}\label{app:baselines_version}

% 画个类似下面的表格 写一下调用的版本号就行

Table~\ref{tab:model_versions} summarizes the specific model versions and API endpoints used for each vision-language model evaluated within the \textsc{WebVIA} framework.

\begin{table*}[t!]
\centering
\small
\renewcommand{\arraystretch}{1.2}
\caption{Versions and official API endpoints for each evaluated vision-language model. }
\begin{tabular}{l l l}
\hline
\textbf{Model} & \textbf{Version} & \textbf{API Endpoint / URL} \\
\hline
Claude-Sonnet-4 & claude-sonnet-4-20250514-thinking & \url{https://www.anthropic.com/api} \\
Claude-Opus-4 & claude-opus-4-20250514-thinking & \url{https://www.anthropic.com/api} \\
Claude-Sonnet-3.7 & claude-3-7-sonnet-20250219-thinking & \url{https://www.anthropic.com/api} \\
GPT-5 & gpt-5-2025-08-07 & \url{https://platform.openai.com/docs/models} \\
o4-mini & o4-mini-2025-04-16 & \url{https://platform.openai.com/docs/models} \\
GPT-4o & gpt-4o-2024-11-20 & \url{https://platform.openai.com/docs/models} \\
Gemini-2.5-pro & gemini-2.5-pro-preview-06-05 & \url{https://ai.google.dev/gemini-api/docs/models} \\
Gemini-2.5-flash & gemini-2.5-flash-preview-05-20 & \url{https://ai.google.dev/gemini-api/docs/models} \\
\hline
\end{tabular}
\label{tab:model_versions}
\end{table*}

\subsection{Prompts for Baseline Models}
\label{app:prompts}

To ensure a fair comparison across all baselines, we designed unified prompt templates for the two main subtasks in the \textsc{WebVIA} framework: (1) \textit{Action Generation} and (2) \textit{UI2Code Translation}. 
Each baseline model (e.g., Claude-Sonnet, GPT-5, Gemini-2.5) was queried using the same textual instructions, with minimal format adaptation to comply with their API requirements. 
Temperature was fixed to $0.0$ for deterministic outputs unless otherwise noted.

\vpara{Action Generation Prompt}
This task evaluates a model’s ability to identify and describe actionable interactive elements given a static webpage representation. Specifically, the model is provided with a webpage screenshot and its corresponding DOM tree and is required to generate a set of valid user actions (e.g., clicks, text inputs, or selections) that can be performed on the interface. The complete instruction template used for this task is illustrated in Figure~\ref{fig:action_generation_prompt}.

\begin{figure*}[t]
\centering
\begin{tcolorbox}[
    colback=gray!2,
    colframe=blue!70!black,
    colbacktitle=blue!10!white,
    title=\textbf{Action Generation Prompt},
    fonttitle=\bfseries,
    coltitle=blue!60!black,
    enhanced,
    sharp corners,
    boxrule=0.4pt,
    left=5pt,
    right=5pt,
    top=5pt,
    bottom=5pt,
    titlerule=0mm,
    title style={left color=blue!8!white, right color=blue!5!white},
]
\small
You are an interactive web assistant. I now want to check whether all interactive buttons on this webpage work properly. For example, if there is a search box on the page, please search with a reasonable query and click confirm, expecting the page to change. Note that if multiple interactive components are almost identical, please select only one of them. For example, if the page has multiple similar items each with an "Edit" button, please choose only once.

\vspace{1mm}
The current page state is part of the detection process. I will send you which components have already been clicked. If you find that an image was clicked before, please focus on what is different in the image I send you this time compared with the previous one. For example, if a new window has popped up, please make sure to only select interactive components in the new part. If you find that the image I send you this time is almost identical to one of the historical ones (for example, all buttons are the same, with only minor text differences), then directly reply with: “All operations on this page are completed”.

\vspace{1mm}
Note: If two interactive buttons are not sequentially related (for example, two separate click buttons on the same page), please include only one in each boxed response, separating them. If they are sequentially related (for example, entering multiple values and then clicking confirm), please put them together in the same boxed response. Wrap your answers in LaTeX using \texttt{\textbackslash boxed\{\}}.  

\vspace{1mm}
\textbf{Action Format:}  
\begin{itemize}
    \item \texttt{click[id]} = click  
    \item \texttt{enter[id][text]} = input text  
    \item \texttt{select[id][text]} = select option  
\end{itemize}
Separate each action with a comma.  

\vspace{1mm}
\textbf{DOM elements clicked previously:}  
\texttt{\{history\_info\_prompt\}}  

\vspace{1mm}
\textbf{Important:} Please return only the answer! Do not include anything extra!  

\vspace{1mm}
\textbf{Page Information:}  
\texttt{\{domtree\}}  
\end{tcolorbox}
\caption{Prompt used for action generation in WebVIA.}
\label{fig:action_generation_prompt}
\end{figure*}

\vpara{Interaction Verification Prompt.} To evaluate whether a predicted interaction leads to a functionally correct state transition, the WebVIA-Agent employs a specialized verification prompt.
Given a sequence of webpage screenshots before and after user actions, the model is required to determine whether the visual and structural changes align with the expected interaction outcome.
This prompt guides the agent to reason about the consistency between DOM transitions and visual differences, distinguishing successful interactions (e.g., modals opening, content updates) from failed or redundant ones.
The complete verification prompt is shown in Figure~\ref{fig:interaction_verification_prompt}.

\begin{figure*}[t]
\centering
\begin{tcolorbox}[
    colback=gray!2,
    colframe=blue!70!black,
    colbacktitle=blue!12!white,
    title=\textbf{Interaction Verification Prompt},
    fonttitle=\bfseries,
    coltitle=blue!50!black,
    enhanced,
    sharp corners,
    boxrule=0.5pt,
    left=5pt,
    right=5pt,
    top=5pt,
    bottom=5pt,
    titlerule=0mm,
    title style={left color=blue!8!white, right color=blue!5!white},
]
\small
You will receive multiple webpage images as part of the verification process for interaction history.  
The multiple webpage images are arranged in chronological order: the last image represents the completion of the interaction, and the first image is the starting image.  
A screenshot is taken after each operation until the final image completes the operation.  

For example, if there are only two images, then only one operation was performed:  
the pre-operation screenshot is the first image, and the post-operation screenshot is the second image.  
If there are four images, then three operations were performed: the pre-operation screenshot is the first image, after the first operation is the second image, after the second operation is the third image, and so on.

\vspace{1mm}
\textbf{Tasks:}
\begin{enumerate}
    \item \textbf{Interaction Consistency Check:}  
    Determine whether the webpage shows changes consistent with the described interactive components after this interaction sequence.  
    For example, clicking “Edit” should open an editing window; entering values and clicking “Save” should persist changes; clicking “Cancel” or “Close” should not.  
    Carefully compare the final and initial images to infer whether the expected modification occurred.  
    Reply “\texttt{Yes}” if changes are functionally correct, or “\texttt{No}” if the images remain largely unchanged.  
    Extract your final answer and place it inside LaTeX \texttt{\textbackslash boxed\{\}}, followed by your reasoning.
    
    \item \textbf{Continuation Check:}  
    Compare the last image with the starting image to determine whether any new significant part has appeared on the webpage.  
    If new interactive content appears and further checking is needed, wrap “Continue” with \texttt{\textbackslash terminate\{Continue\}}.  
    If no new meaningful change is observed (e.g., no new section or trivial modifications), wrap “Complete” with \texttt{\textbackslash terminate\{Complete\}}.  
    Both boxed answers and termination tags must be output, with detailed reasoning provided afterward.
\end{enumerate}

\vspace{1mm}
\textbf{Interactive Action / Component Name:} \\
\texttt{<interact\_element\_names>}
\end{tcolorbox}
\caption{Prompt used for interaction verification during the WebVIA-Agent.}
\label{fig:interaction_verification_prompt}
\end{figure*}

\subsection{Interactive Code Generation Evaluation Prompts}\label{app:ui2code_prompts}

To rigorously evaluate the functionality and interactivity of the generated code, we design a three-stage prompting protocol that aligns with the validation module described in Section~\ref{sec:part3}. 
Each stage corresponds to a distinct phase of task-oriented execution and enables consistent benchmarking of action reasoning, process tracking, and outcome verification.

\vpara{(1) Initial Action Selection Prompt.}  
At the beginning of each evaluation episode, the model receives a predefined task description (e.g., “search for an item,” “fill out and submit a form,” or “navigate to the contact page”) along with the initial webpage screenshot and DOM tree. As shown in Figure~\ref{fig:initial_action_selection}, the following prompt is used to request the model’s first interaction decision.

\begin{figure*}[h]
\centering
\begin{tcolorbox}[
    colback=gray!2,
    colframe=blue!70!black,
    colbacktitle=blue!12!white,
    title=\textbf{Action Selection Prompt},
    fonttitle=\bfseries,
    coltitle=blue!50!black,
    enhanced,
    sharp corners,
    boxrule=0.5pt,
    left=5pt,
    right=5pt,
    top=5pt,
    bottom=5pt,
    titlerule=0mm,
    title style={left color=blue!8!white, right color=blue!5!white},
]
\small
You are an interactive web assistant. I now want to check whether certain interactive buttons on this webpage are working properly.  
I will give you several tasks. You need to read the current page and select an action sequence for each task.  
If you think the current page content is insufficient to complete a task, please only select the interactive components on the page that can accomplish part of the task.

\vspace{2mm}
Wrap each of your interactive components in LaTeX \texttt{\textbackslash boxed\{\}}.  
Action format: \texttt{click[id] = click}, \texttt{enter[id][text] = input}, \texttt{select[id][text] = select option}.  
Separate each action with a comma. Please note, \texttt{id} refers to the identifier of this component in the DOM tree.

\vspace{2mm}
At the same time, before each \texttt{\textbackslash boxed\{\}}, write \texttt{\textbackslash task\{<task name>\}},  
and after each \texttt{\textbackslash boxed\{\}}, write \texttt{\textbackslash state\{Complete\}} or \texttt{\textbackslash state\{Continue\}}.

\vspace{2mm}
\textbf{Task list:} \texttt{\{str(tasks)\}} \\
\textbf{Page information:} \texttt{\{domtree\}}
\end{tcolorbox}
\caption{Prompt used in the Validation Module of the WebVIA framework to guide task-specific action selection.}
\label{fig:initial_action_selection}
\end{figure*}

\vpara{(2) Process-State Action Prompt.}  
After executing the initial interaction, the webpage transitions into a new state.  
For each task branch (typically five per task), the model observes the updated screenshot and DOM tree corresponding to the current state and must decide the subsequent action.  
If the task is incomplete, the following process prompt is used (See Figure~\ref{fig:action_execution_prompt}).

\begin{figure*}[h]
\centering
\begin{tcolorbox}[
    colback=gray!2,
    colframe=blue!70!black,
    colbacktitle=blue!12!white,
    title=\textbf{Process-State Action Prompt},
    fonttitle=\bfseries,
    coltitle=blue!50!black,
    enhanced,
    sharp corners,
    boxrule=0.5pt,
    left=5pt,
    right=5pt,
    top=5pt,
    bottom=5pt,
    titlerule=0mm,
    title style={left color=blue!8!white, right color=blue!5!white},
]
\small
You are an interactive web assistant. I now want you to complete a task.  
You are currently in the detection process. Please read which buttons have already been clicked on the historical pages.  
Only select the buttons on the page that can actually be clicked.  
You should focus only on completing one task. Read the current page and select an ongoing action sequence for the current task.  
If you think the current page content is insufficient to complete the task, please only select the interactive components on the page that can accomplish part of the task.

\vspace{2mm}
Wrap your interactive components with a LaTeX \texttt{\textbackslash boxed\{\}}.  
Action format: \texttt{click[id] = click}, \texttt{enter[id][text] = input}, \texttt{select[id][text] = select option}.  
Separate each action with a comma.

At the same time, before each \texttt{\textbackslash boxed\{\}}, write \texttt{\textbackslash task\{<task name>\}},  
and after each \texttt{\textbackslash boxed\{\}}, write \texttt{\textbackslash state\{Complete\}} or \texttt{\textbackslash state\{Continue\}}.

\vspace{2mm}
\textbf{Task content:} \texttt{\{task\_text\}} \\
\textbf{Page information:} \texttt{\{domtree\}}
\end{tcolorbox}
\caption{Action execution prompt used in the Validation Module of the WebVIA framework for determining actionable components and task progress on interactive webpages.}
\label{fig:action_execution_prompt}
\end{figure*}

\vpara{(3) Task Completion Verification Prompt.}  
Once a task branch reaches termination, we verify whether the task goal has been successfully accomplished.  
The model receives the full textual task description and the sequence of screenshots collected during its execution.  As shown in Figure~\ref{fig:task_completion_prompt}, the following prompt is used for task-level verification.

\begin{figure*}[t]
\centering
\begin{tcolorbox}[
    colback=gray!2,
    colframe=blue!70!black,
    colbacktitle=blue!12!white,
    title=\textbf{Task Completion Verification Prompt},
    fonttitle=\bfseries,
    coltitle=blue!50!black,
    enhanced,
    sharp corners,
    boxrule=0.5pt,
    left=5pt,
    right=5pt,
    top=5pt,
    bottom=5pt,
    titlerule=0mm,
    title style={left color=blue!8!white, right color=blue!5!white},
]
\small
Provide all the screenshots along this path (in chronological order). \\ 
\textbf{Task:} \texttt{\{task\_text\}}

\vspace{2mm}
Please determine whether the expected webpage changes for this task have been completed. \\ 
\textbf{Important:} Each task name corresponds to a single interactive button or operation.  
For example, “New” only represents opening the new page, not saving.  
Thus, you only need to verify whether the new page can be opened.  
Only if the task explicitly specifies “New - Input ... - Save” do you need to confirm the saving step.  
Similarly, “Delete” only refers to opening the delete dialog, while “Delete - Confirm Delete” represents two operations—only in the latter case should you check whether the deletion was actually completed.

\vspace{2mm}
If the operation has been successfully completed, respond with \texttt{\textbackslash boxed\{Yes\}};  
if not completed, respond with \texttt{\textbackslash boxed\{No\}}.  
Afterward, briefly explain your reasoning.
\end{tcolorbox}
\caption{Prompt used by Validation Module in the WebVIA framework to determine whether a task has been successfully completed based on sequential webpage screenshots.}
\label{fig:task_completion_prompt}
\end{figure*}

\subsection{Demo Cases for Exploration}\label{app:exploration}

To demonstrate the versatility and robustness of the WebVIA exploration process, we present qualitative examples of the agent’s behavior in both \textit{real-world} and \textit{synthetic} webpage environments. These cases highlight how WebVIA effectively handles complex UI layouts, multi-step operations, and dynamic visual feedback during autonomous exploration.

\begin{figure*}[h]
    \centering
    \includegraphics[width=1\linewidth]{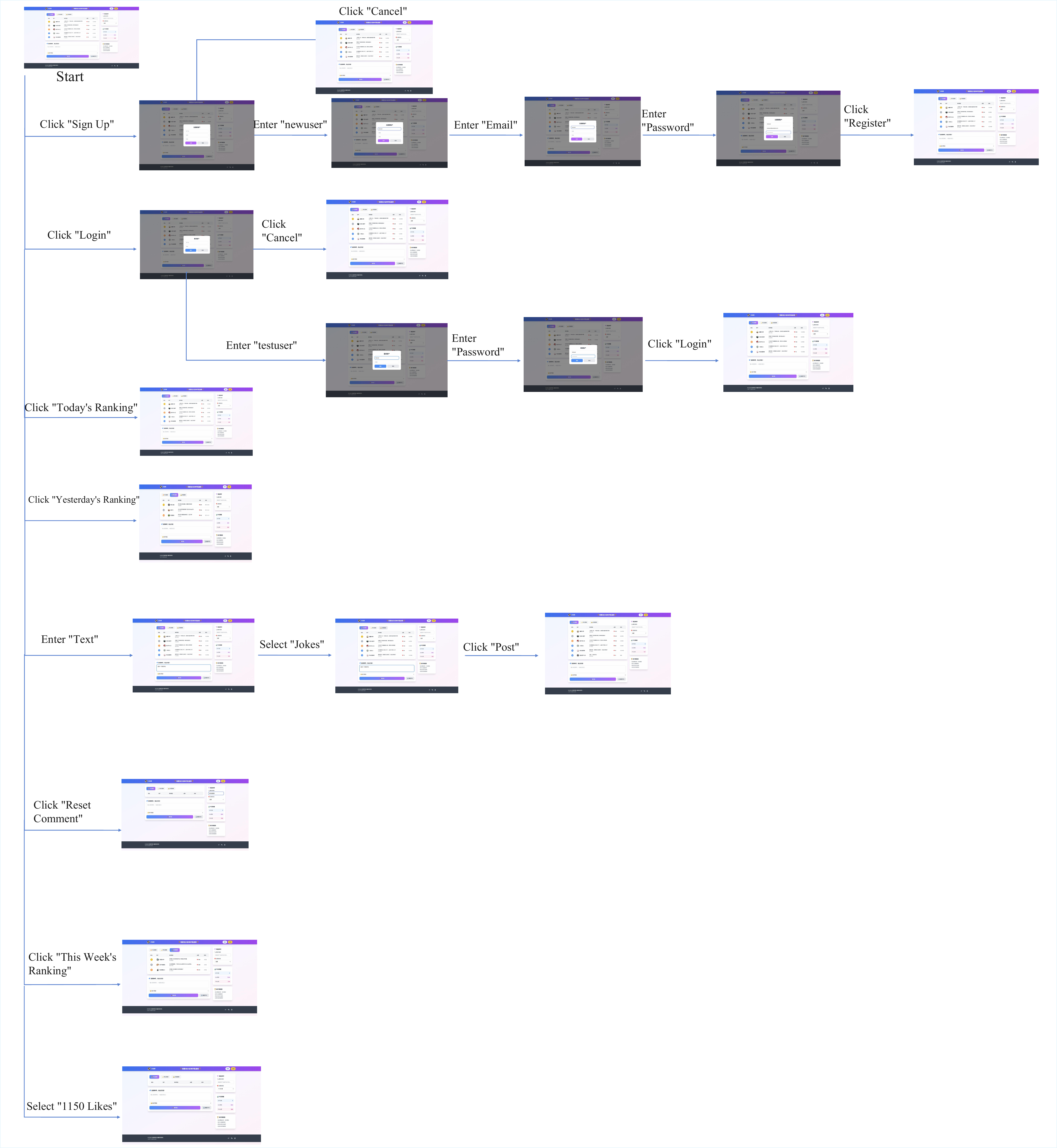}
    \caption{Exploration results of the \textbf{WebVIA-Agent} on synthesized web environments.}
    \label{fig:demo-1-syn}
\end{figure*}

\begin{figure*}[h]
    \centering
    \includegraphics[width=1\linewidth]{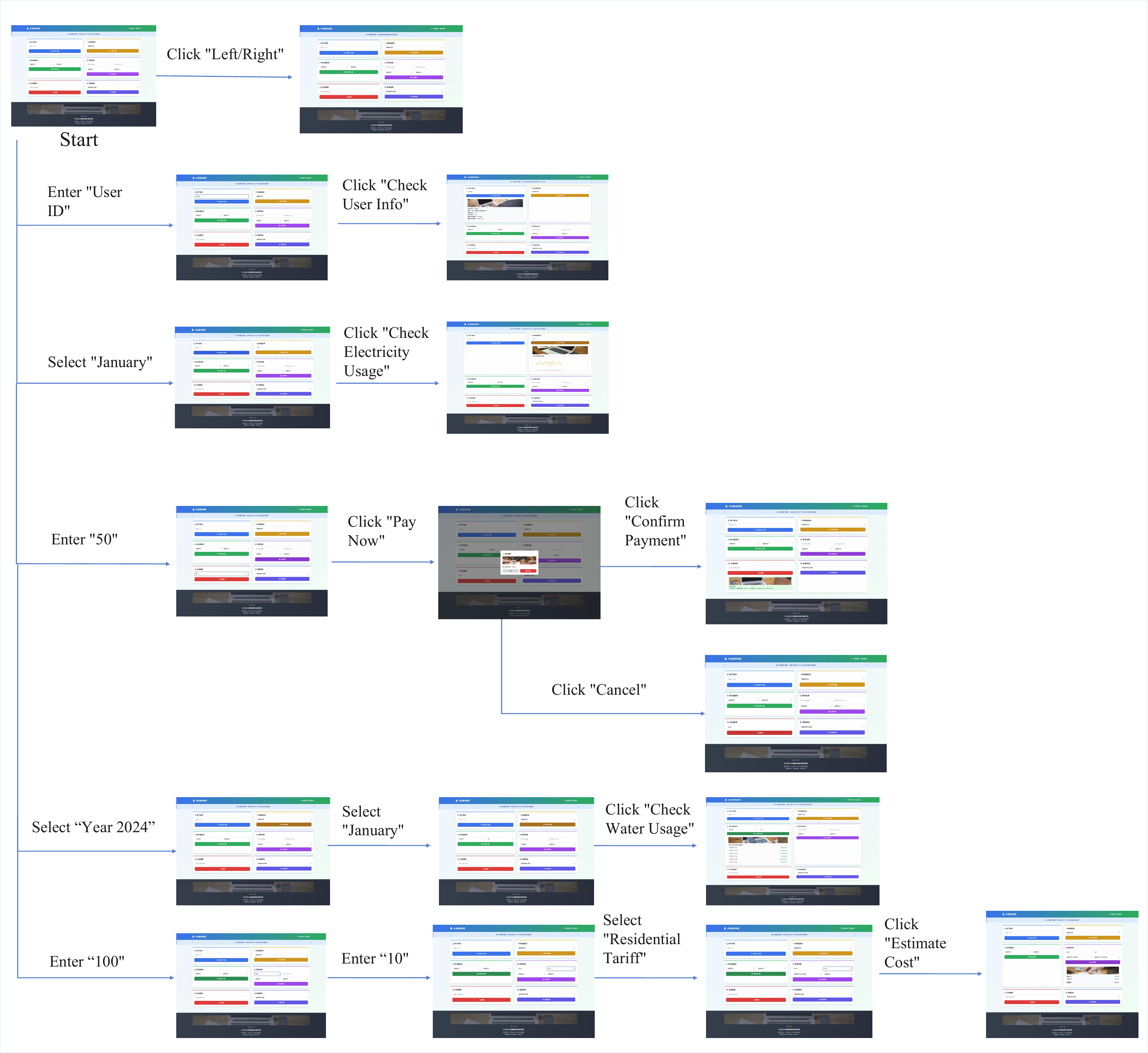}
    \caption{Exploration results of the \textbf{WebVIA-Agent} on synthesized web environments.}
    \label{fig:demo-2-syn}
\end{figure*}

\begin{figure*}[h]
    \centering
    \includegraphics[width=1\linewidth]{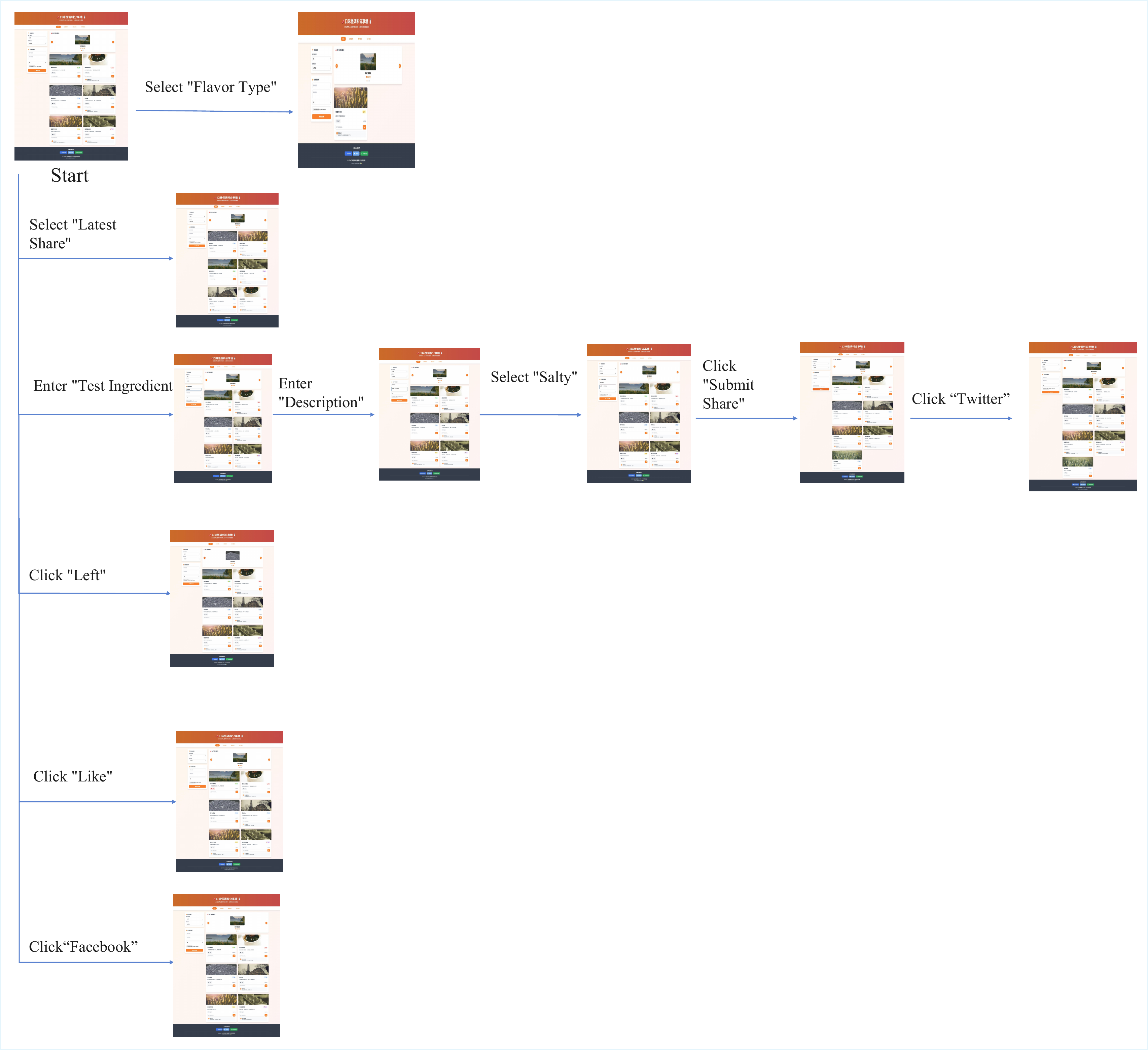}
    \caption{Exploration results of the \textbf{WebVIA-Agent} on synthesized web environments.}
    \label{fig:demo-3-syn}
\end{figure*}

\begin{figure*}[h]
    \centering
    \includegraphics[width=1\linewidth]{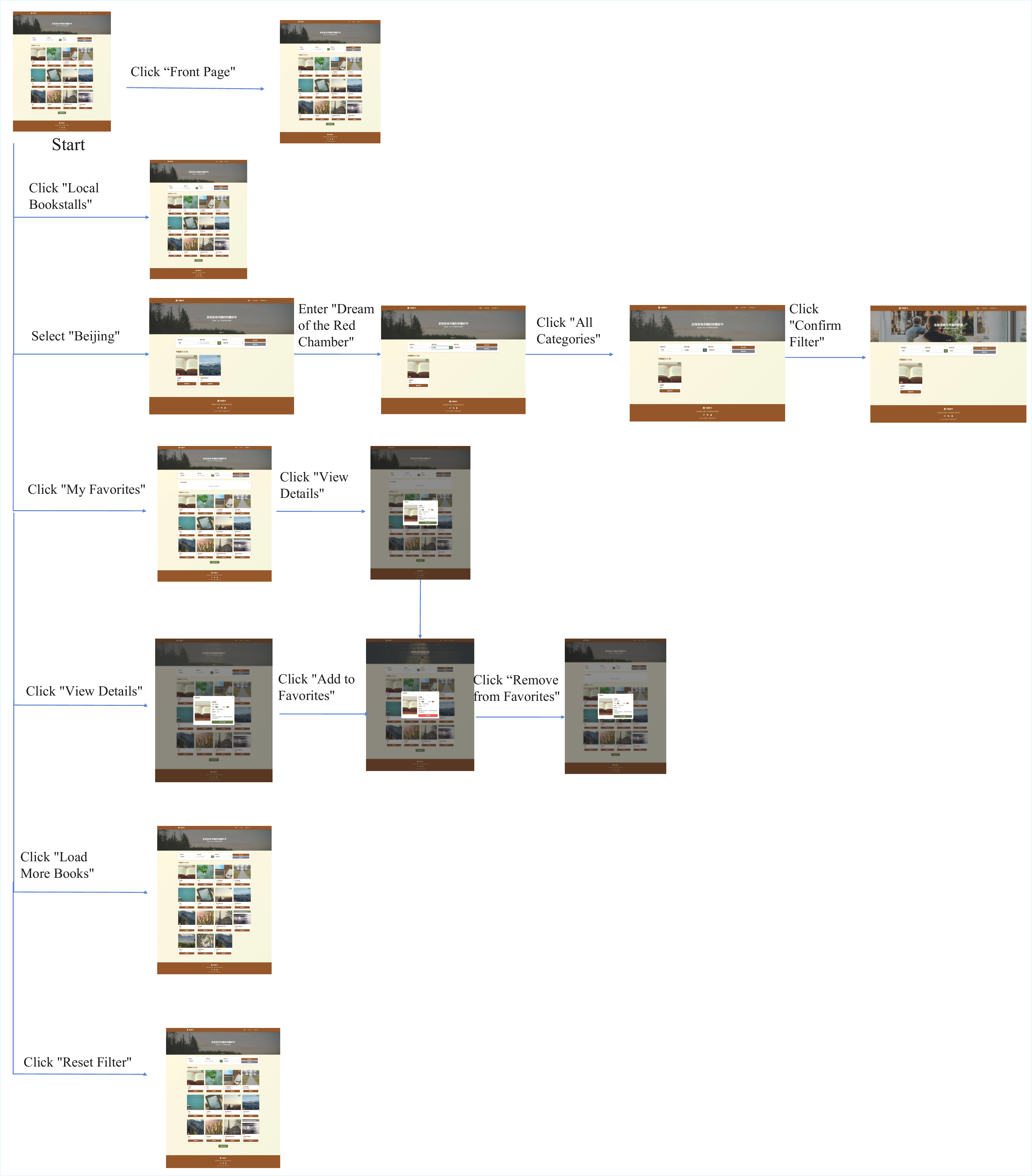}
    \caption{Exploration results of the \textbf{WebVIA-Agent} on synthesized web environments.}
    \label{fig:demo-4-syn}
\end{figure*}

\vpara{Synthetic Webpage Exploration.} 
Figures~\ref{fig:demo-1-syn}–\ref{fig:demo-4-syn} demonstrate the exploration trajectory of WebVIA-Agent on the synthetic webpages. Each figure corresponds to a distinct interaction scenario generated within our procedural environment. 
The agent autonomously identifies visible interactive components such as buttons, forms, and dropdowns, and performs multi-step actions to manipulate the webpage state. 
Across these examples, WebVIA-Agent demonstrates its ability to perceive layout hierarchies, maintain consistency between visual and structural states, and accurately capture interaction outcomes.
These exploration traces form the foundation for downstream UI2Code synthesis and interaction verification.

\begin{figure*}
    \centering
    \includegraphics[width=\linewidth]{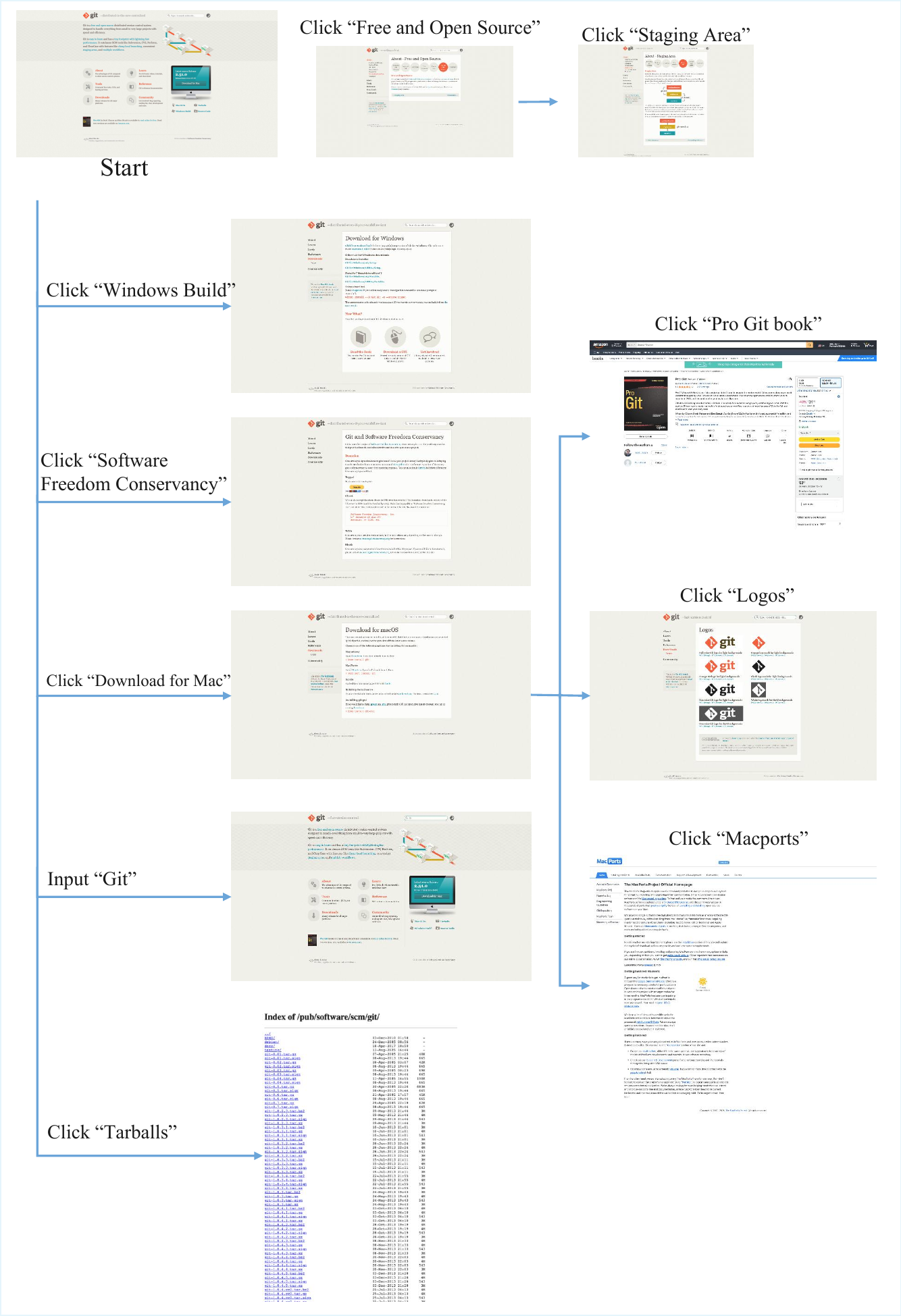}
    \caption{Exploration results of the \textbf{WebVIA-Agent} on real-world web environments.}
    \label{fig:rw_agent_1}
\end{figure*}

\begin{figure*}
    \centering
    \includegraphics[width=\linewidth]{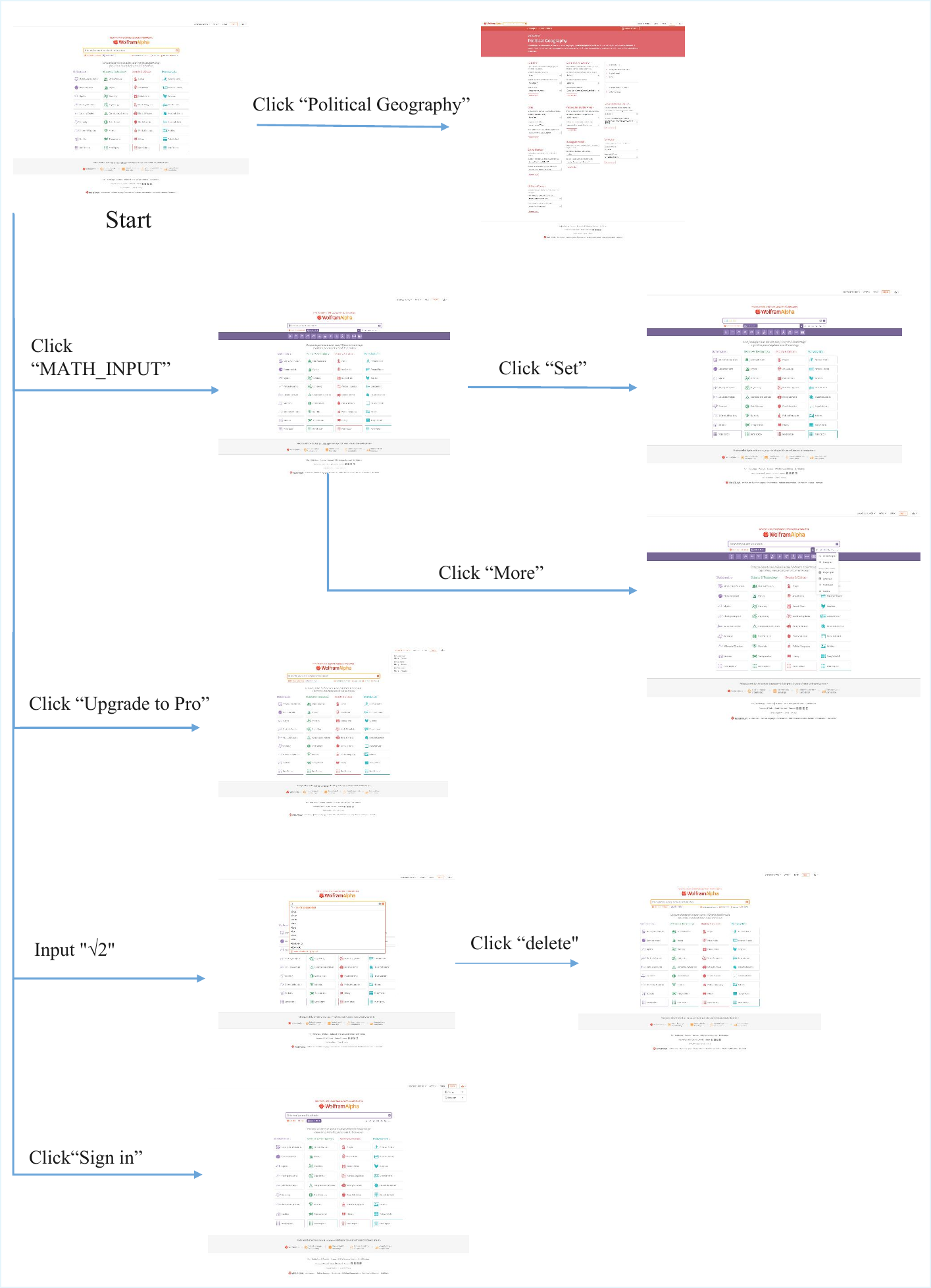}
    \caption{Exploration results of the \textbf{WebVIA-Agent} on real-world web environments.}
    \label{fig:rw_agent_2}
\end{figure*}

\begin{figure*}
    \centering
    \includegraphics[width=\linewidth]{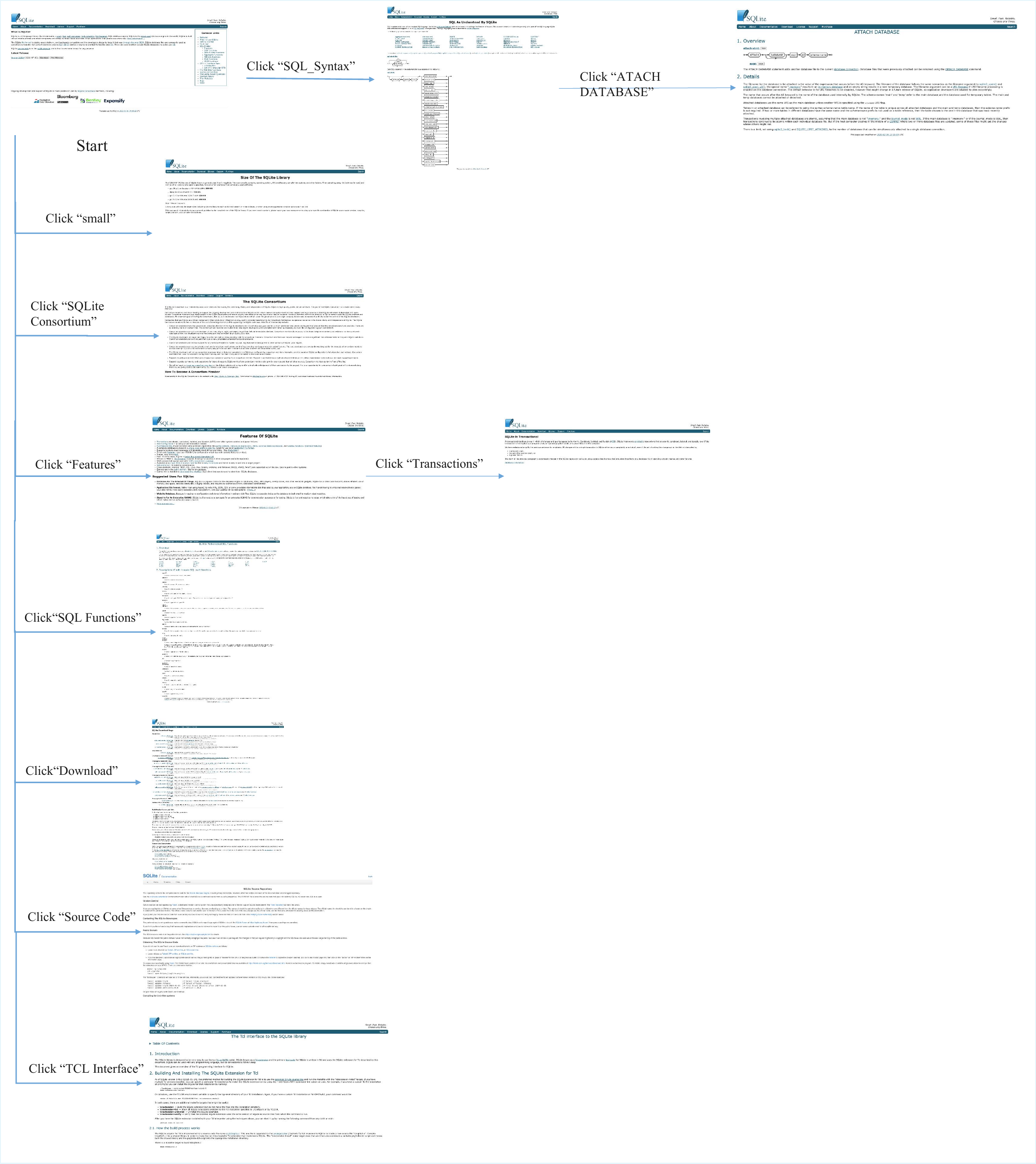}
    \caption{Exploration results of the \textbf{WebVIA-Agent} on real-world web environments.}
    \label{fig:rw_agent_3}
\end{figure*}

\begin{figure*}
    \centering
    \includegraphics[width=\linewidth]{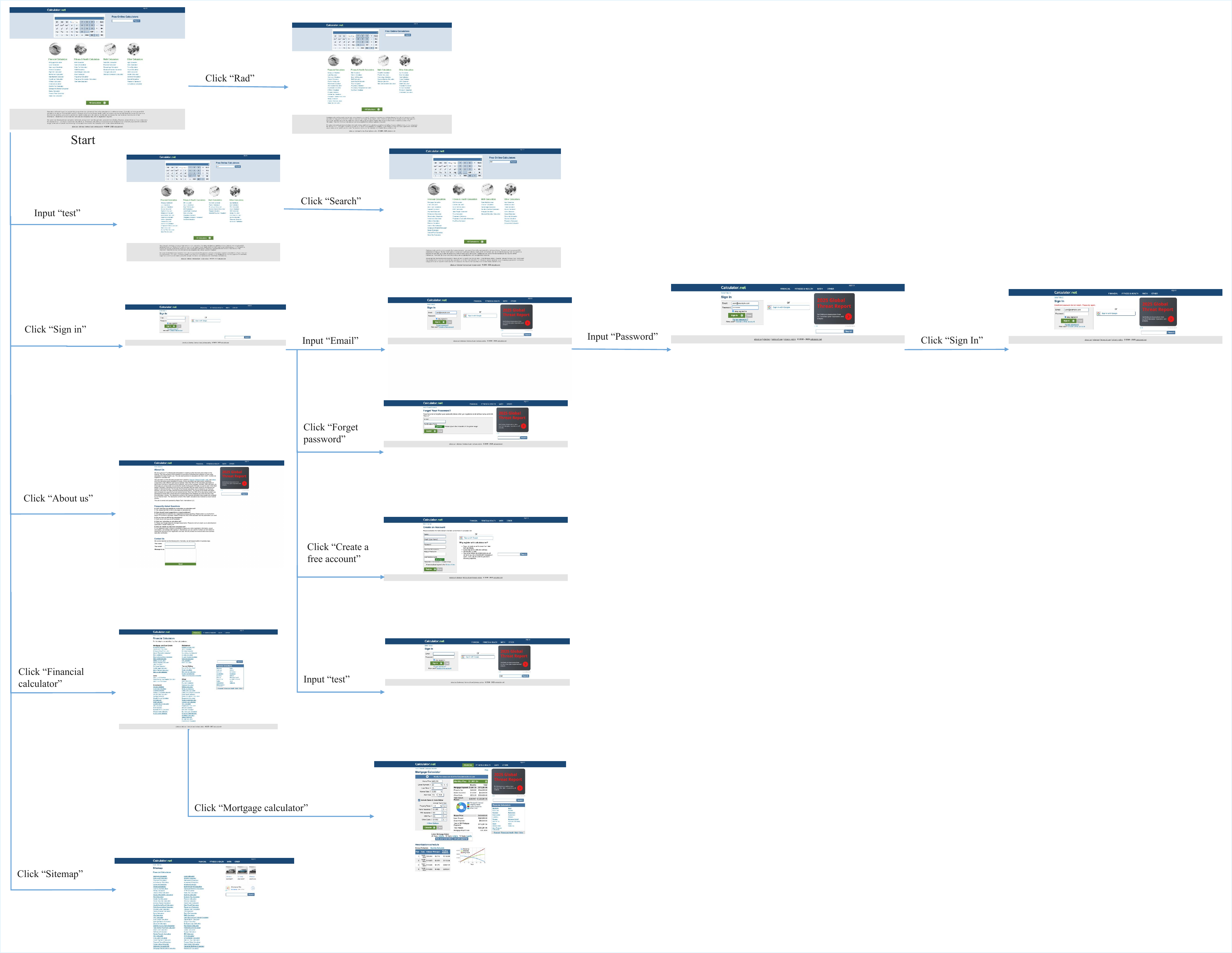}
    \caption{Exploration results of the \textbf{WebVIA-Agent} on real-world web environments.}
    \label{fig:rw_agent_4}
\end{figure*}

\begin{figure*}
    \centering
    \includegraphics[width=\linewidth]{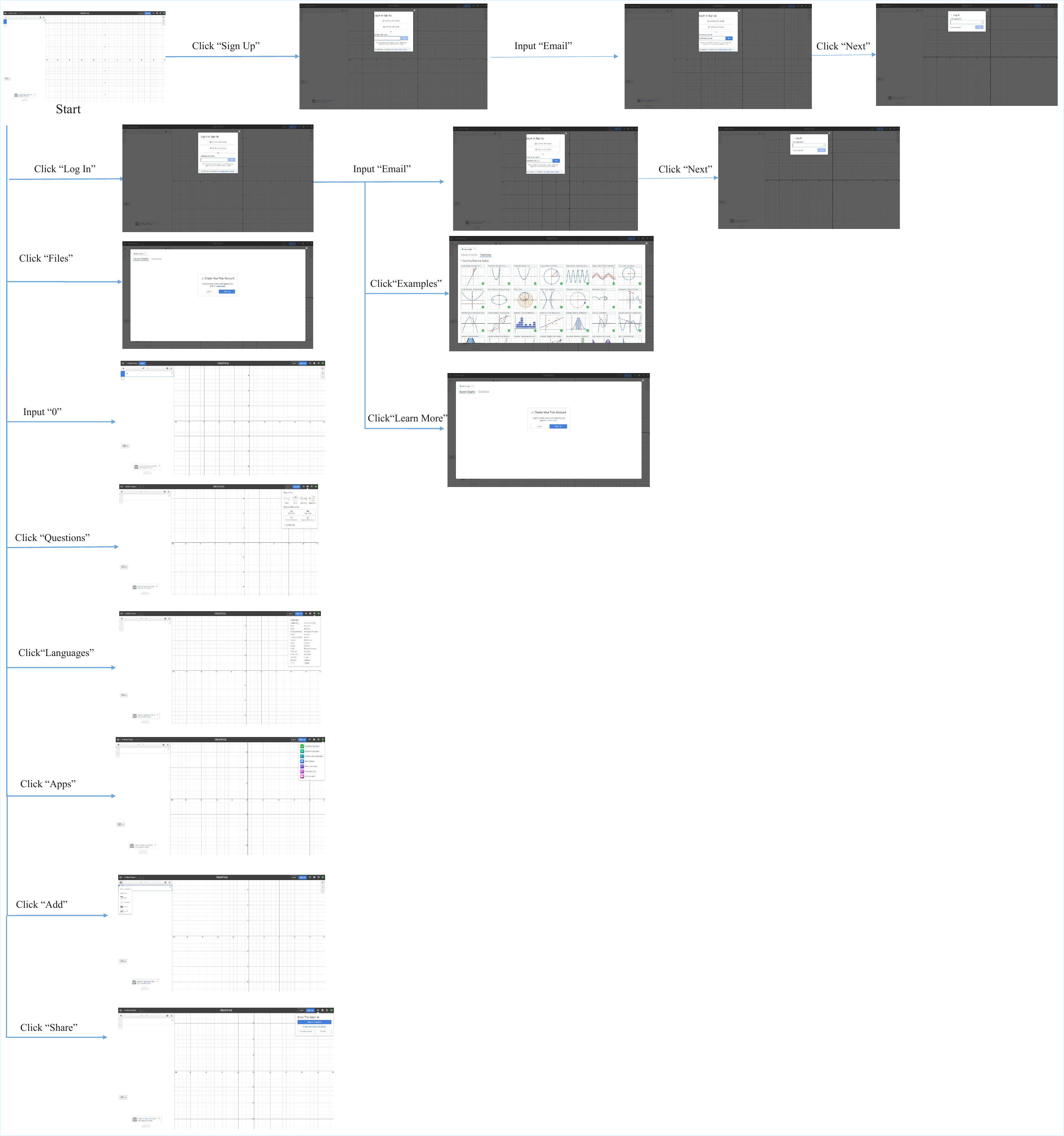}
    \caption{Exploration results of the \textbf{WebVIA-Agent} on real-world web environments.}
    \label{fig:rw_agent_5}
\end{figure*}

\vpara{Real-World Webpage Exploration.} 
Figures~\ref{fig:rw_agent_1}–\ref{fig:rw_agent_5} present qualitative demonstrations of \textbf{WebVIA-Agent} exploring real-world webpages collected from open-access sites. 
Notably, although the agent is trained exclusively within our synthetic environment, it generalizes effectively to complex real webpages \textit{without any additional fine-tuning}.
It can accurately identify functional UI components—such as navigation bars, search boxes, and modal dialogs—and execute multi-step interactions involving both visual reasoning and structural understanding. 
During exploration, the agent maintains alignment between rendered screenshots and DOM hierarchies, correctly detecting dynamic transitions.  
These results demonstrate WebVIA-Agent’s strong zero-shot generalization ability from procedurally generated environments to real web interfaces, validating the robustness and transferability of its visual–structural reasoning process.

\subsection{Demo Cases for Interactive Code Generation}\label{app:ui2code_generation}

\begin{figure*}
    \centering
    \includegraphics[width=\linewidth]{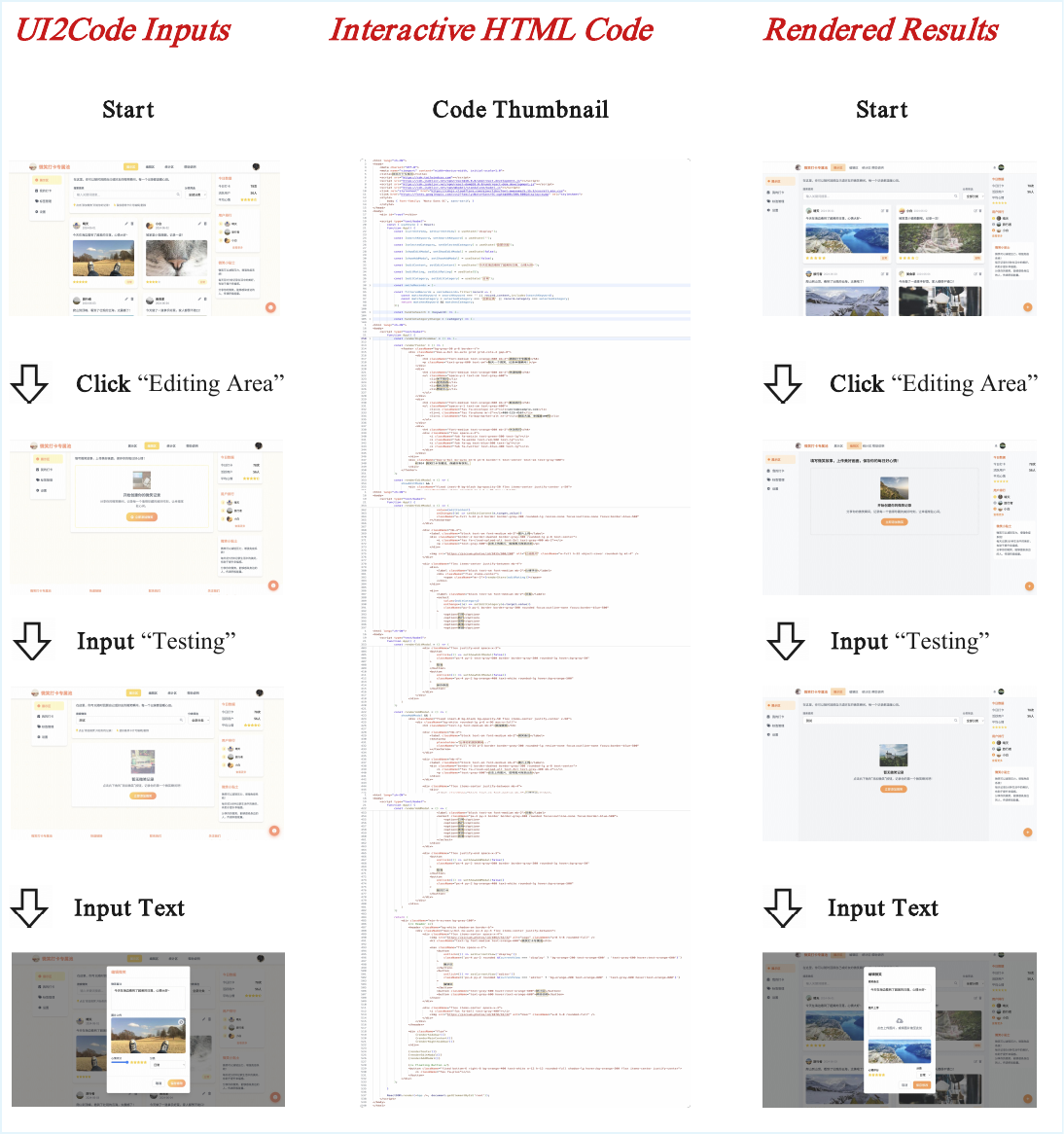}
    \caption{Rendered UI2Code demo for WebVIA-UI2Code-GLM}
    \label{fig:ui2code_synth1}
\end{figure*}

\begin{figure*}
    \centering
    \includegraphics[width=\linewidth]{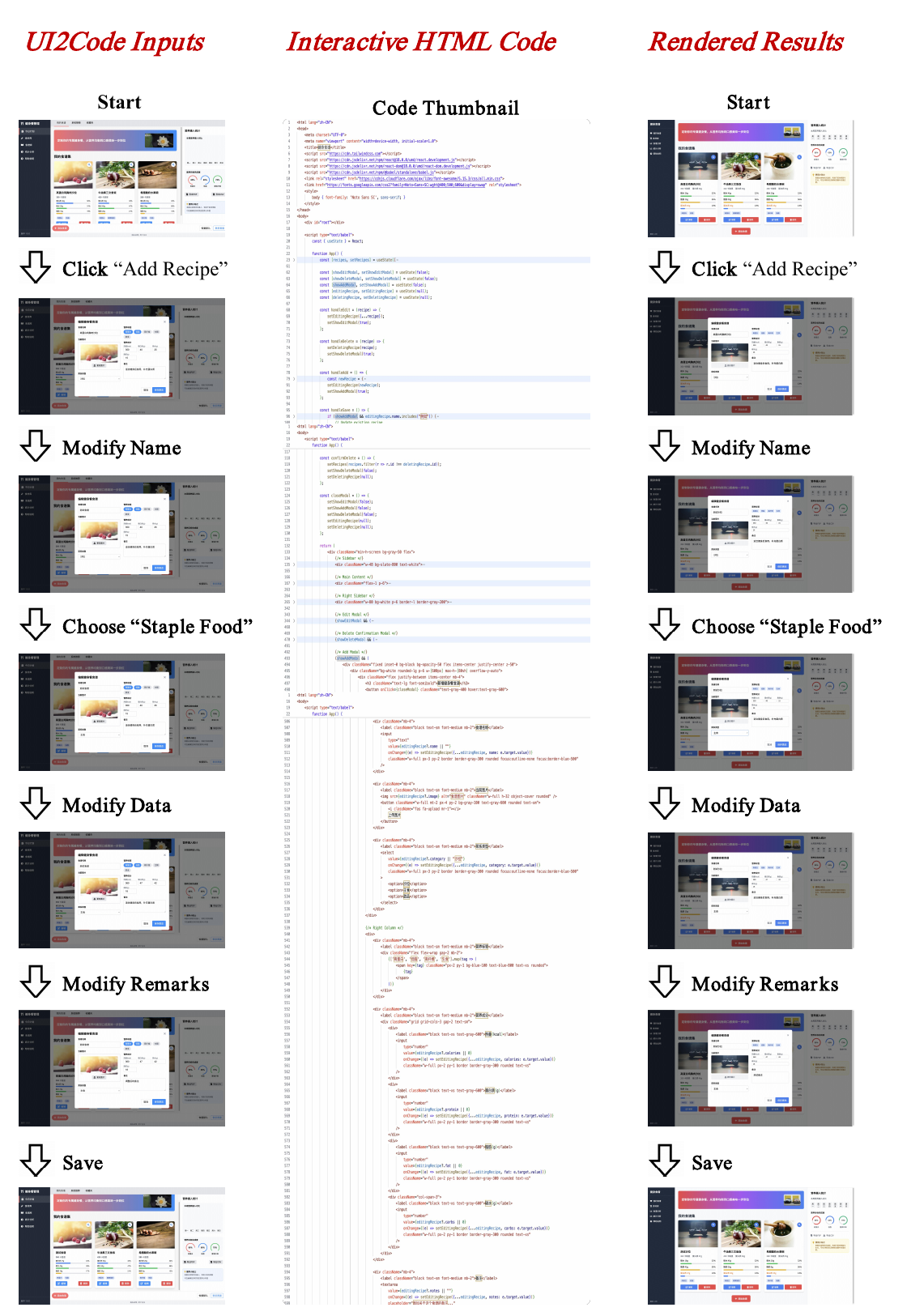}
    \caption{Rendered UI2Code demo for WebVIA-UI2Code-GLM}
    \label{fig:ui2code_synth2}
\end{figure*}

\begin{figure*}
    \centering
    \includegraphics[width=\linewidth]{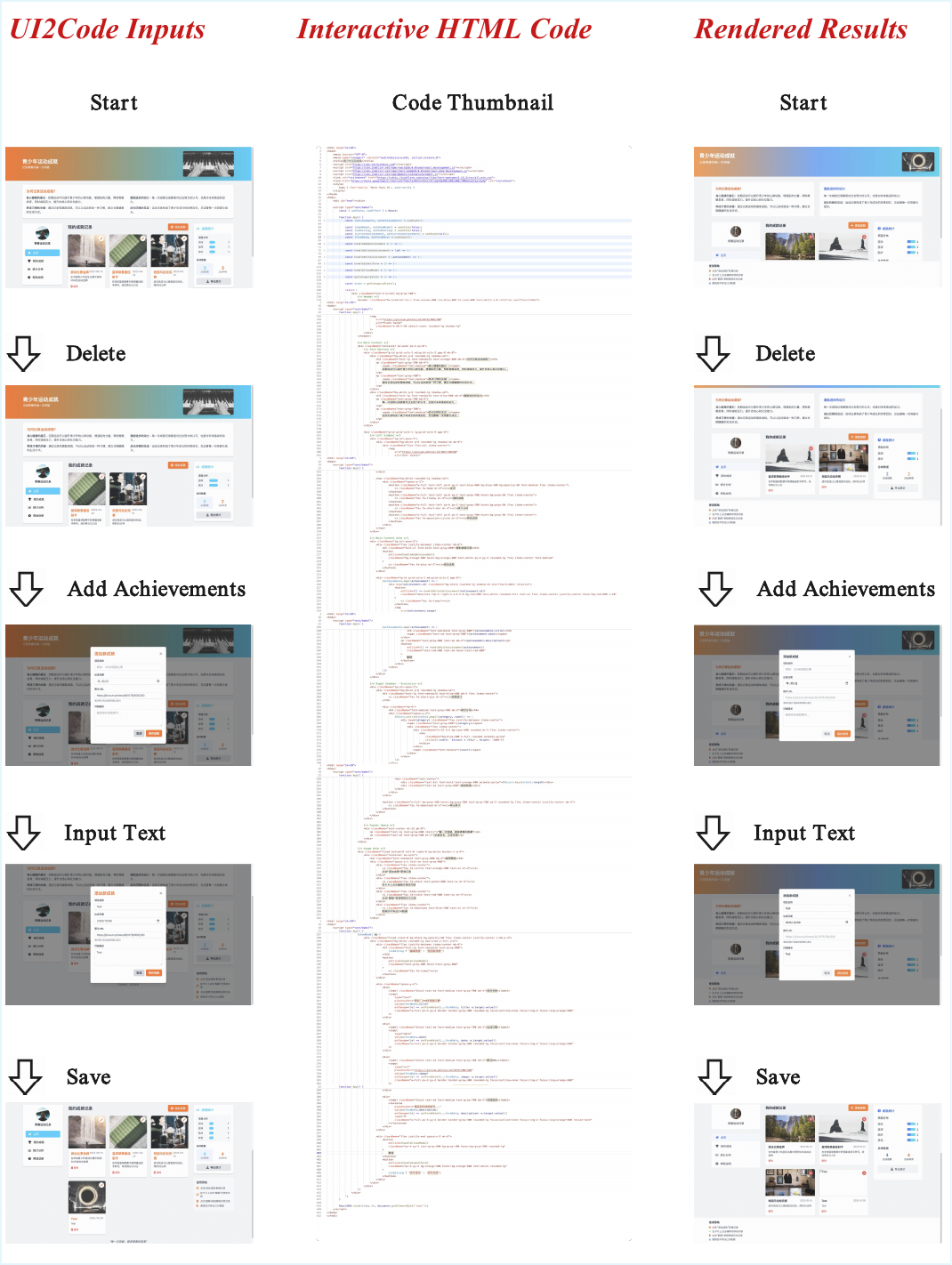}
    \caption{Rendered UI2Code demo for WebVIA-UI2Code-GLM}
    \label{fig:ui2code_synth3}
\end{figure*}

\begin{figure*}
    \centering
    \includegraphics[width=\linewidth]{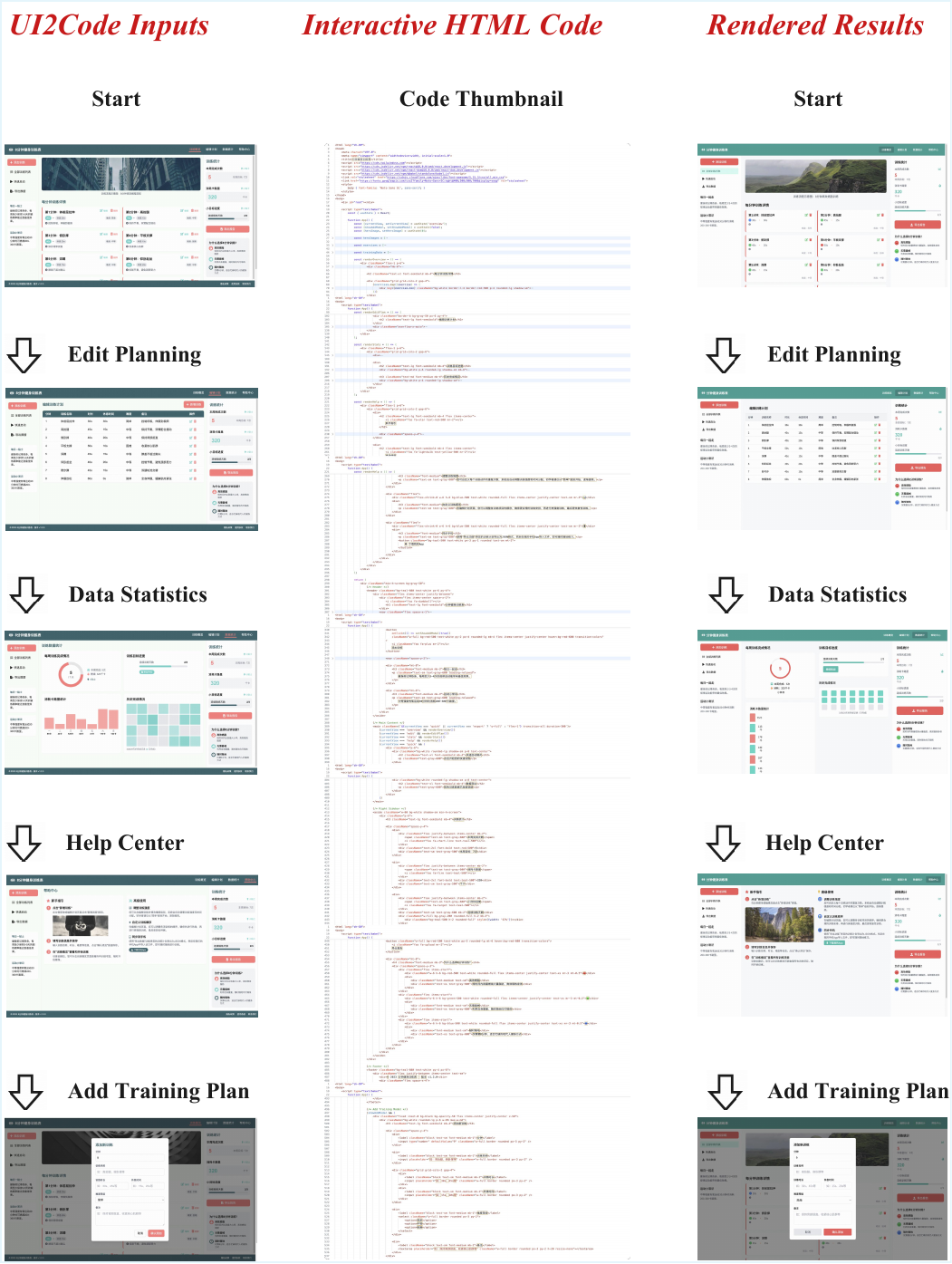}
    \caption{Rendered UI2Code demo for WebVIA-UI2Code-GLM}
    \label{fig:ui2code_synth4}
\end{figure*}

\begin{figure*}
    \centering
    \includegraphics[width=\linewidth]{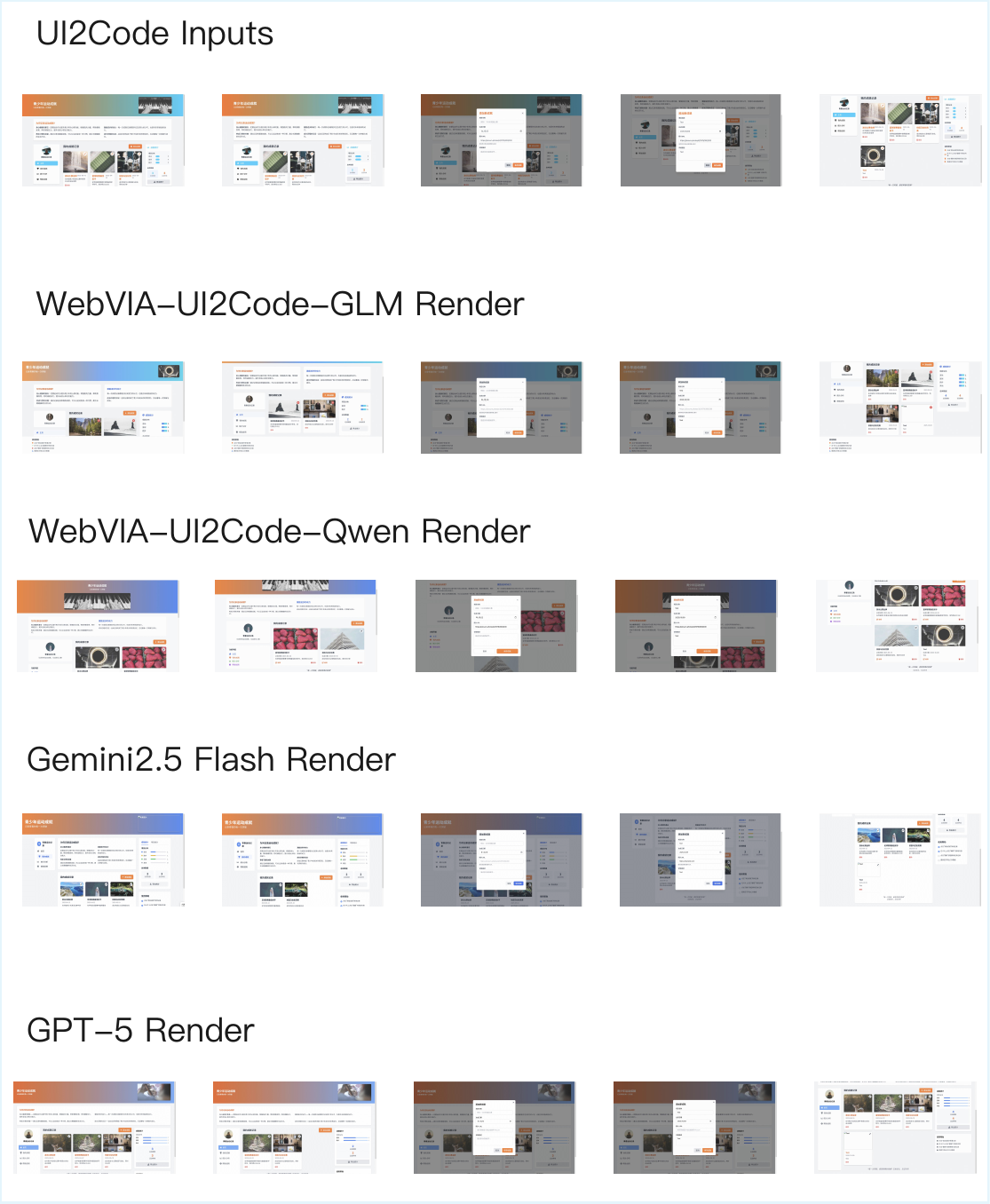}
    \caption{Comparison of WebVIA-UI2Code and baseline renders on the same interaction trace.}
    \label{fig:ui2code_synth4_compare}
\end{figure*}

\begin{figure*}
    \centering
    \includegraphics[width=\linewidth]{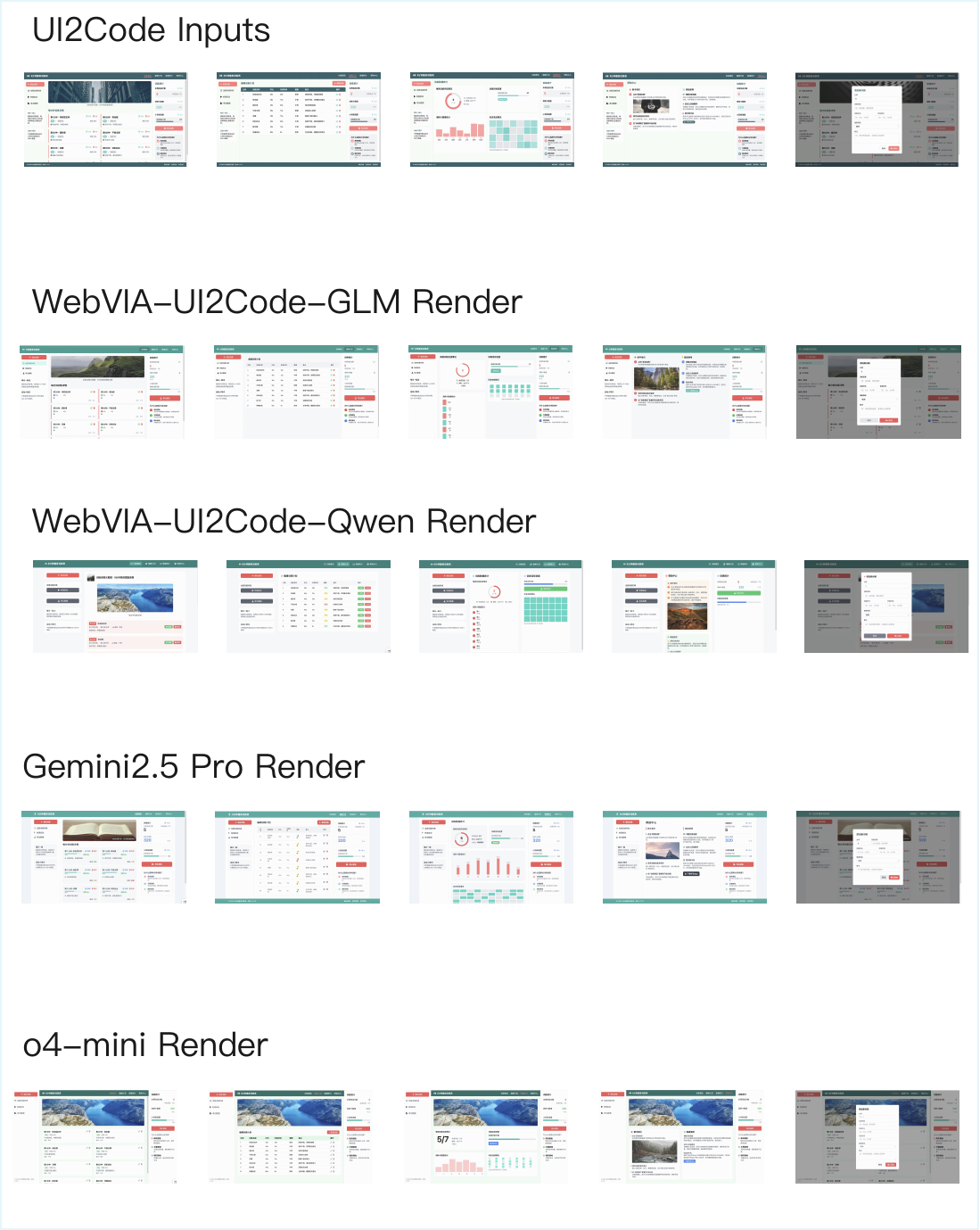}
    \caption{Comparison of WebVIA-UI2Code and baseline renders on the same interaction trace.}
    \label{fig:ui2code_synth1_compare}
\end{figure*}

To further illustrate the capabilities of \textbf{WebVIA-UI2Code}, we showcase qualitative results of interactive UI-to-Code generation. Figures~\ref{fig:ui2code_synth1}–\ref{fig:ui2code_synth4} demonstrate WebVIA-UI2Code-GLM performing interactive code generation in procedurally synthesized environments. Each synthetic webpage is automatically composed of diverse UI components such as navigation bars, cards, modals, and dropdowns, each associated with predefined interaction logic. 
The model observes sequential webpage states during user–interface interactions and generates executable React + Tailwind code that faithfully reproduces both the visual layout and dynamic behaviors observed in the interaction flow.  Furthermore, we conduct a comparative study across multiple models, where each model receives the same sequence of \textit{multiple UI screenshots} as input and is tasked with generating interactive HTML code.  As shown in Figures~\ref{fig:ui2code_synth1_compare}–\ref{fig:ui2code_synth4_compare}, WebVIA-UI2Code-GLM consistently produces structurally complete and functionally executable webpages, while baseline models often fail to maintain state consistency or omit interaction logic.

\begin{figure*}
    \centering
    \includegraphics[width=\linewidth]{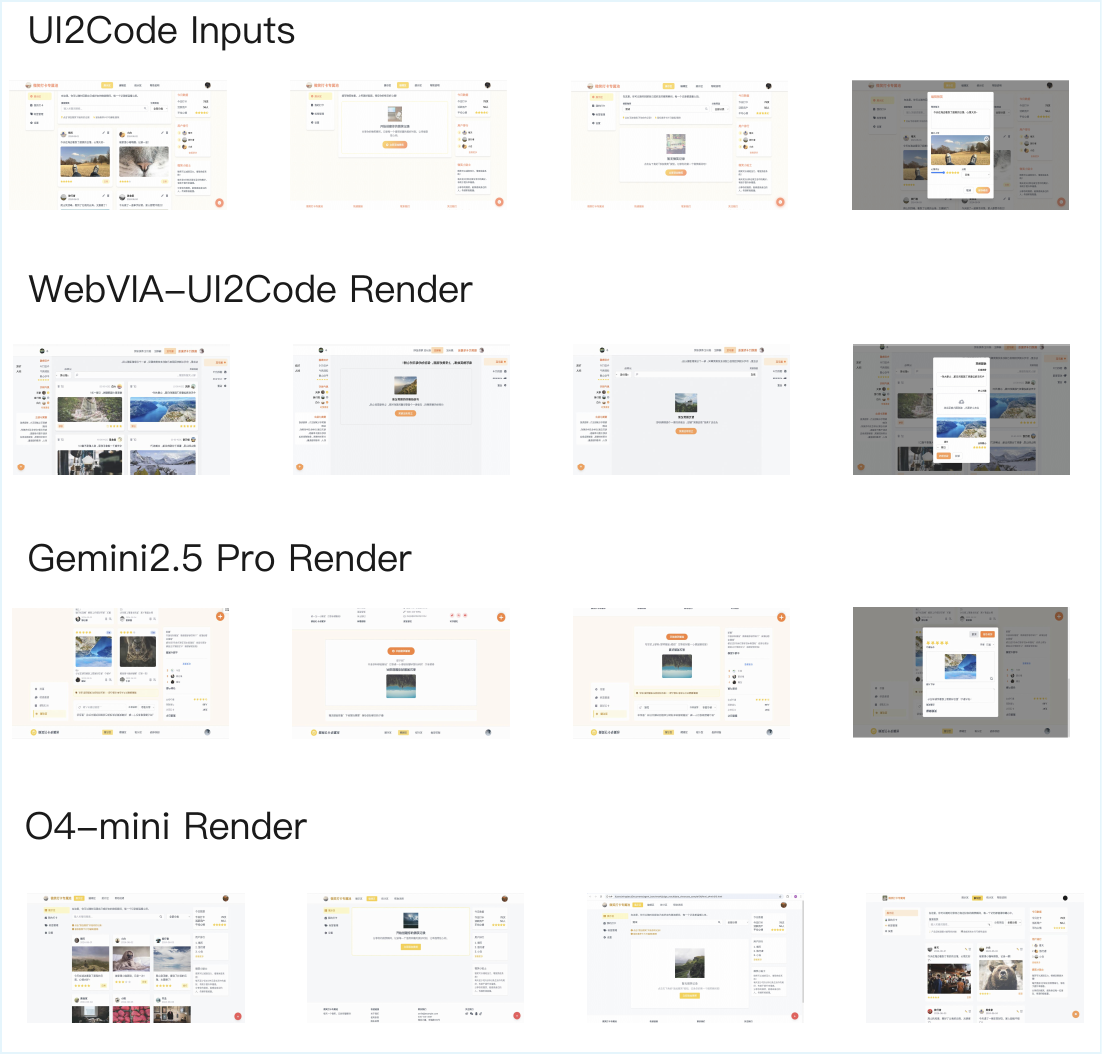}
    \caption{Comparison of WebVIA-UI2Code and baseline renders on the same interaction trace.}
    \label{fig:ui2code_synth2_compare}
\end{figure*}

\begin{figure*}
    \centering
    \includegraphics[width=\linewidth]{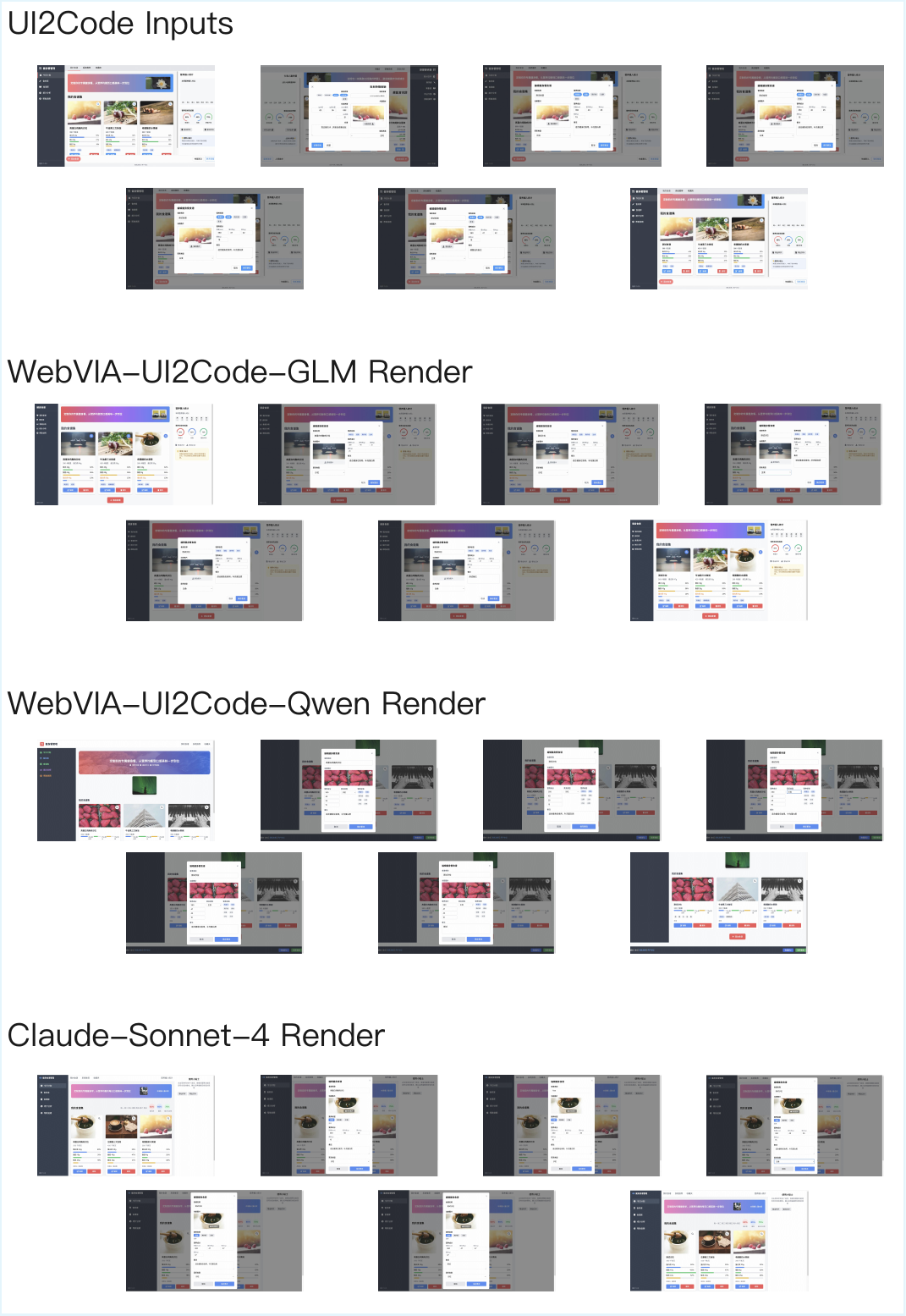}
    \caption{Comparison of WebVIA-UI2Code and baseline renders on the same interaction trace.}
    \label{fig:ui2code_synth3_compare}
\end{figure*}

\end{document}